\documentclass[twocolumn]{aastex62}
\usepackage{graphicx,subfigure,amsmath,amsfonts, amssymb,footnote,color,epstopdf,ulem}
\usepackage[bookmarks=false]{hyperref}
\usepackage{apjfonts}
\hypersetup{
  colorlinks= true, %Colours links instead of ugly boxes
  urlcolor  = blue, %Colour for external hyperlinks
  linkcolor = red, %Colour of internal links
  citecolor = blue %Colour of citations
} % makes stupid green boxes go away, paper now looks more like it will when published!

\shorttitle{}
\shortauthors{Lacy et al.}

\begin{document}


%\title{A ground-based extragalactic survey at high angular resolution using multiconjugate adaptive optics and deep radio observations.} 
%\title{A ground-based, high angular resolution survey of star formation and AGN activity in three deep fields.} 

%\title{A subarcsecond Gemini near-infrared view of massive galaxies at z > 1:  the compactness of AGN hosts and the discovery of a candidate triple AGN}
%\title{A subarcsecond Gemini near-infrared view of massive galaxies at z > 1:  Compact AGN hosts and a candidate triple AGN}
\title{A subarcsecond near-infrared view of massive galaxies at z > 1 with Gemini Multiconjugate Adaptive Optics}


\author{M.\ Lacy} 
\affiliation{National Radio Astronomy Observatory, 520 Edgemont Road, Charlottesville, VA 22903, USA}
\author{K.\ Nyland}
\affiliation{National Radio Astronomy Observatory, 520 Edgemont Road, Charlottesville, VA 22903, USA}
\author{M.\ Mao}
\affiliation{Jodrell Bank Centre for Astrophysics, Alan Turing Building, School of Physics and Astronomy, The University of Manchester, Oxford Road, Manchester, M13 9PL, UK}
\author{P.\ Jagannathan},
\affiliation{National Radio Astronomy Observatory, 1003, Lopezville Road, Socorro, NM 87801, USA}
\author{J.\ Pforr}
\affiliation{ESA/ESTEC Keplerlaan1, 2201 AZ, Noordwijk, The Netherlands.}
\author{S.E.\ Ridgway}
\affiliation{National Optical Astronomy Observatory, 950 North Cherry Avenue, Tucson, AZ, 85719, USA}
\author{J.\ Afonso}
\affiliation{Instituto de Astrof\'{i}sica e Ci\^{e}ncias do Espa\c co, Universidade de Lisboa, OAL, Tapada da Ajuda, PT1349-018 Lisboa, Portugal}
\affiliation{Departamento de F\'{i}sica, Faculdade de Ci\^{e}ncias, Universidade de Lisboa, Edif\'{i}cio C8, Campo Grande, PT1749-016 Lisbon, Portugal}
\author{D.\ Farrah}
\affiliation{Department of Physics, Virginia Tech, Blacksburg, VA 24061, USA}
\author{P.\ Guarnieri}
\affiliation{Institute of Cosmology and Gravitation, Dennis Sciama Building, Burnaby Road, Portsmouth PO1 3FX, UK}
\author{E.\ Gonzales-Solares}
\affiliation{Institute of Astronomy, Madingley Rd, Cambridge CB3 0HA, UK}
\author{M.J.\ Jarvis},
\affiliation{Astrophysics, Department of Physics, Keble Road, Oxford OX1 3RH, UK}
\affiliation{Department of Physics and Astronomy, University of the Western Cape, Private Bag X17, 7535, Bellville, Cape Town, South Africa}
\author{C.\ Maraston}
\affiliation{Institute of Cosmology and Gravitation, Dennis Sciama Building, Burnaby Road, Portsmouth PO1 3FX, UK}
\author{D.M.\ Nielsen}
\affiliation{Dept. of Astronomy, University of Wisconsin - Madison, 475 N. Charter Street, Madison, WI, 53706, USA}
\author{A.O.\ Petric}
\affiliation{CFHT Corporation, 65-1238 Mamalahoa Hwy, Kamuela, HI 96743, USA}
\author{A.\ Sajina}
\affiliation{Department of Physics and Astronomy, Tufts University, 212 College Avenue, Medford, MA 02155, USA}
\author{J.A.\ Surace}
\affiliation{Spitzer Science Center, California Institute of Technology, Pasadena, CA 91125, USA}
\author{M.\ Vaccari}
\affiliation{Department of Physics and Astronomy, University of the Western Cape, Private Bag X17, 7535, Bellville, Cape Town, South Africa}
\affiliation{INAF - Istituto di Radioastronomia, via Gobetti 101, 40129 Bologna, Italy}

\begin{abstract}
We present images taken using the Gemini South Adaptive Optics Imager (GSAOI) with the Gemini Multiconjugate Adaptive Optics System (GeMS)  in three 2 arcmin$^2$ fields in the {\em Spitzer} Extragalactic Representative Volume Survey. These GeMS/GSAOI observations are among the first $\approx 0\farcs1$ resolution data in the near-infrared spanning extragalactic fields exceeding $1\farcm5$ in size. 
We use these data to estimate galaxy sizes, obtaining results similar to those from studies with the {\em Hubble Space Telescope}, though we find a higher fraction of compact star forming galaxies at $z>2$.
To disentangle the star-forming galaxies from active galactic nuclei (AGN), we use multiwavelength data from surveys in the optical and infrared, including far-infrared data from {\it Herschel}, as well as new radio continuum data from the Australia Telescope Compact Array and Very Large Array.  We identify ultraluminous infrared galaxies (ULIRGs) at $z \sim 1-3$, which consist of a combination of pure starburst galaxies and Active Galactic Nuclei (AGN)/starburst composites. The ULIRGs show signs of recent merger activity, such as highly disturbed morphologies and include a rare candidate triple AGN. We find that AGN tend to reside in hosts with smaller scale sizes than purely star-forming galaxies of similar infrared luminosity. Our observations demonstrate the potential for MCAO to complement the deeper galaxy surveys to be made with the {\em James Webb Space Telescope}.
\end{abstract}


\keywords{galaxies:evolution -- galaxies:active -- galaxies: interactions -- galaxies: photometry -- techniques: high angular resolution}

\section{Introduction}

Our current understanding of the evolution of massive galaxies invokes their formation at 
%high redshift ($z\sim 2-7$).  
high redshift.  
%Studies show that such galaxies are significantly more compact than galaxies of comparable mass at low redshift (e.g.\ Daddi et al.\ 2005; Buitrago et al.\ 2008; Newman et al.\ 2012; Patel et al.\ 2017), and that they probably form ``inside out'' as highly obscured objects whose stars are concentrated at the centers of gas disks (van Dokkum et al.\ 2015). How these galaxies expand their scale sizes to produce the population seen today is also matter of active research, though it probably involves accretion of material via minor mergers (e.g.\ Hilz et al. 2013). However, there remain many unanswered questions, such as the role of further star formation and major mergers. Also unclear} is why the star formation rate of a typical galaxy close to the knee in the galaxy mass function is much higher at redshifts $z\sim 1-3$ than in the current epoch (e.g.\ Rodighiero et al.\ 2014, Speagle et al. 2014). 
Studies show that such galaxies are significantly more compact than galaxies of comparable mass at low redshift (e.g. Daddi et al. 2005; Buitrago et al. 2008; Newman et al. 2012; Petty et al.\ 2013; Patel et al. 2017).  The origin of this population of compact, massive galaxies at high-z has been attributed to a phase of rapid gas accretion at early times (e.g.\ Dekel \& Burkert 2014), or major merger activity (e.g.\ Wuyts et al.\ 2010).  The gas from these processes sinks toward the center of the galaxy, thus spurring bulge formation, heavily obscured star formation, and massive black hole growth. Subsequent growth via dry, stellar mergers is believed to be responsible for increasing galaxy sizes to produce the observed population seen today (Cappellari 2016).  However, many unanswered questions remain, such as the redshift evolution of basic galaxies properties such as mass and size, the role of major mergers and the relative importance of different star formation quenching mechanisms in making massive galaxies passive (Wellons et al.\ 2015; Furlong et al. 2017;
Pandya et al.\ 2017).  Also unclear is why the star formation rate of a typical galaxy close to the knee in the galaxy mass function is much higher at redshifts $z\sim 1-3$ than in the current epoch (e.g.\ Rodighiero et al.\ 2014, Speagle et al.\ 2014). 

Determining the physical mechanisms by which massive galaxies evolve into the objects we see today requires imaging high-$z$ galaxies on sub-kpc scales. Imaging in the rest-frame optical/near-infrared, longward of the Balmer break, where the stellar population is dominated by the older stars, is particularly valuable compared to imaging the rest-frame UV, where the light is dominated by young stars and morphologies can be strongly affected by dust. Such imaging can be used to measure both the changing distribution of galaxy sizes as a function of redshift (e.g.\ van der Wel et al.\ 2014) and the frequency of interactions and mergers. Furthermore, by combining near-infrared imaging of the stellar light with high resolution radio continuum imaging (which pinpoints the regions of star formation or nuclear activity in these systems) we can build up a much more complete picture of the nature of the galaxies. This is particularly valuable in dusty star-forming systems, where the peak of star formation activity may be offset from the peak of the visible stellar light (Rujopakarn et al.\ 2016, 2018).
%light, {\bf or be distributed around the a core of older stars in the ``inside out'' formation model.}

Measurement of the morphologies of galaxies on sub-arcsecond scales has been dominated by observations made with the {\em Hubble Space Telescope (HST)}. Adaptive Optics (AO) from the ground with large aperture  telescopes is capable of delivering better image resolution at near-infrared wavelengths, but studies of field galaxies have, until recently, been restricted to small ($\approx 30^{\prime \prime}$ diameter) patches of sky near individual guide stars (e.g. Glassman, Larkin \& Lafreni\`{e}re  2002). The advent of Multi-Conjugate Adaptive Optics (MCAO) allows larger fields (up to $\approx 1\farcm5$ in diameter) to be imaged by correcting multiple layers of the atmosphere, probed by multiple guide stars. This overcomes two limitations of conventional AO: 1) the limitation of the $\sim 30^{\prime \prime}$ radius isoplanatic patch over which correction from a single guide star is effective, and 2) the ``cone effect'' from laser guide stars, whereby the atmospheric turbulence probed by a single laser guide star is not the same as that from an arriving wavefront from a distant star (Tallon \& Foy 1990). The Gemini Multiconjugate adaptive optics System (GeMS) on the Gemini South 8m telescope (Rigaut et al.\ 2014; Neichel et al.\ 2014) uses a five laser guide star and a natural guide star constellation of between one and three stars (with the best wide-field correction made with three stars) to achieve a consistent point spread function (PSF) over a significantly wider ($\sim 1\farcm5$) field of view. Schirmer et al.\ (2015) demonstrate the effectiveness of this technique through imaging of the Hubble Frontier Fields MACS0416.1-2403 and Abell 2744, and Sweet et al.\ (2017) highlight the technique through a study of  cluster galaxies at $z=1$.

By obtaining subarcsecond-resolution imaging over a wide field of view, we can survey the sizes and morphologies of both the faint galaxy population and also extreme galaxies at high redshift, such as the hosts of Active Galactic Nuclei (AGN) and Ultraluminous Infrared Galaxies (ULIRGs). As AO works well in the near-infrared $K$-band, at longer wavelengths than {\em HST} (whose longest wavelength filter on an operational instrument is in $H$-band), AO is very suitable for investigating dusty galaxies detected in the infrared (e.g.\ Melbourne et al.\ 2011; Perna et al.\ 2015). Furthermore, the resolution available in these fields is sufficient to comfortably resolve the majority of the galaxy population over a wide range of redshifts. 
%This, combined with the relatively constant PSF across the field of view, means that MCAO data resemble current space-based data.

Despite the recent advances in adaptive optics technology, the current use of GeMS is restricted to asterisms having stars brighter than $R\approx 15$ (depending on observing conditions), ideally consisting of three stars in an approximately equilateral triangle, and within an $\approx 2^{\arcmin}$ field of view (Neichel et al.\ 2014). Such asterisms are rare far from the Galactic Plane (we find $\sim 1$~deg$^{2}$) and are even less commonly found in well-studied small area deep extragalactic fields, which are typically picked to avoid bright stars. Fortunately, the new generation of deep, wide area ($>>1$~deg$^2$) extragalactic surveys, designed to study the evolution of galaxies over a wide range in environment, can also complement MCAO facilities by both containing suitable asterisms, and having the multi-wavelength coverage needed to obtain photometric redshifts and star formation rates.
%for rare, luminous objects.} 
The {\it Spitzer} Extragalactic Representative Volume Survey (SERVS; Mauduit et al.\ 2012) and associated VISTA Deep Extragalactic Observations (VIDEO) survey (Jarvis et al.\ 2013) provide 12~deg$^2$ of deep near-infrared observations in seven bands from $0.9-4.5\mu$m, enough area to find several such asterisms. We therefore took the opportunity to select asterisms in this survey suitable for observations with GeMS.

Deep, high resolution imaging in the radio enabled by the new generation of wide-band correlators on radio interferometers has allowed a comparable transformation in the ability to image the faint radio emission from star forming galaxies at $z\sim 1-3$. These observations have sufficient sensitivity to resolve emission on subarcsecond scales (e.g.\ Murphy et al.\ 2017). In cases 
%{\bf where high brightness temperature radio emission can be isolated from surrounding diffuse emission} 
where a radio excess above the maximum level expected from star formation is identified, these observations can distinguish emission due to star formation from that due to AGN (e.g.\ Rujopakarn et al.\ 2016).

In this paper, we present results from a pilot study consisting of observations of the fields of three asterisms with GeMS and the GSAOI near-infrared imager (Table \ref{tab:gemsobs}). We combine the GeMs data with the multi-wavelength data in the survey fields (see Vaccari 2015) to obtain photometric redshifts and galaxy morphologies. We also present deep radio surveys in the two fields that contained the highest numbers of AGN and Herschel sources. The radio surveys were made at arcsecond or better resolution, with the aim of firmly identifying the host galaxies (or host galaxy components in a merging system) responsible for the bulk of the emission related to the AGNs and starbursts. We also discuss how ground-based MCAO observations can be used to complement the deeper observations that will be made with the {\em James Webb Space Telescope (JWST)}. We assume a cosmology with $H_0=70$~km~s$^{-1}$~Mpc$^{-1}$, $\Omega_M=0.3$
and $\Omega_{\Lambda}=0.7$.

\section{Observations}

\subsection{Observed fields}

The three asterisms observed are drawn from three different SERVS fields: ELAIS-S1 %(centered on $00^{\rm h}$:37$^{\rm m}$:48$^{\rm s}$ -44$^{\circ}$:00$^{'}$), XMM-LSS (centered on 02$^{\rm h}$:20$^{\rm m}$:00$^{\rm s}$ -04$^{\circ}$:48$^{'}$) and CDFS (centered on 03$^{\rm h}$:32$^{\rm m}$:19$^{\rm s}$ -28$^{\circ}$:06$^{'}$). 
(centered on R.A.(J2000) $=$ 00:37:48, decl.(J2000) $=-$44:00), XMM-LSS (centered on R.A.(J2000) $=$ 02:20:00, decl.(J2000) $=-$04:48) and CDFS (centered on R.A.(J2000) $=$ 03:32:19, decl.(J2000) $=-$ 28:06). 
All SERVS fields were observed to the same depth (20~min) during the post-cryogenic {\em Spitzer} mission in the 3.6 and $4.5\,\mu$m bands, with the observational scheme described in Mauduit et al.\ (2012), reaching a $5\sigma$ depth of $\approx 2\,\mu$Jy in each band. 

These fields have been the subjects of extensive multiwavelength studies, and include data from the optical through the infrared. Of particular relevance to this paper is the HerMES survey (Oliver et al.\ 2012) with the {\em Herschel} telescope (Pilbratt et al.\ 2010). The {\em Herschel}/SPIRE observations in the three fields differed in exposure time; however, they are all confusion limited at noise levels of $\approx 6$~mJy at 250 and 350~$\mu$m, and $\approx 10$~mJy at $500\,\mu$m. 
In all three fields we used near-infrared $Z, Y, J, H$ and $K_{\rm s}$ data from the VIDEO survey (Jarvis et al.\ 2012) and data from the 
%SWIRE {\it Spitzer} Legacy survey in the 5.8, 8.0, 24 and 70~$\mu$m bands (Lonsdale et al.\ 2003). 
SIRTF Wide-area Infrared Extragalactic Legacy Survey (SWIRE; Lonsdale et al.\ 2003) in the 5.8, 8.0, 24 and 70~$\mu$m bands.  

The source of the optical data depended on the field, in ES1 we used data from our own $i$-band survey (described in Section \ref{sec:servs_i} below) and the preliminary release of the Dark Energy Survey (DES; Dark Energy Survey Collaboration et al.\ 2016), in XMM-LSS we used data from the Canada-France-Hawaii Telescope Legacy Survey (CFHTLS; Gwyn 2012), and in CDFS we used ancillary optical data from the SWIRE survey. Our use of archival datasets, including approximate depths, is summarized in Tables \ref{tab:common} and \ref{tab:optical}.

\subsection{Gemini MCAO Observations}
  
We selected asterisms in the SERVS fields that were as close as possible to the ideal equilateral triangle configuration that yields a near-uniform PSF across the field (Figures \ref{fig:es1c}, \ref{fig:cdfsc} and \ref{fig:xmmc2}, with the PSFs shown in Figure \ref{fig:psfs}). 
%Asterisms like these are rare,
%with approximately one per square degree suitable for GEMS in its current configuration, i.e.\ needing guide
%stars brighter than $R\approx 15$ (depending on conditions). 
%(A planned upgrade to GeMS will see this limit reduced to $R\approx 17$.) 
The three fields were observed in Gemini program GS-2013B-Q14. ES1-C is in the SERVS  ELAIS-S1 field, XMM-C2 in the XMM-LSS field and CDFS-C in the CDFS field. Observations are listed in Table \ref{tab:gemsobs} and further details of the observational conditions for adaptive optics are listed in Table \ref{tab:aoobs}. Typical Strehl ratios (the ratio of the measured PSF peak to that obtained considering only diffraction; see Section 3.9) in $K$-band for the ES1-C and CDFS-C observations were $10-15$\%, comparable to the values seen in commissioning (Neichel et al.\ 2014). The conditions for the XMM-C2 observations were significantly worse, with Strehl ratio $\approx 5$\%. Nevertheless the observations resulted in a better FWHM than obtainable from non-AO observations. Each field is $\approx 2~$arcmin$^2$ in area.

We chose the $K$-band since it is both likely to have the highest Strehl ratio, and it samples the rest-frame optical at $z\sim 3$, yet cannot be observed with {\em HST}. Observations were obtained through the $K^{\prime}$ filter (central wavelength 2.11~$\mu$m), which we judged to be the best compromise between bandwidth and low thermal noise. We used 150-second exposures, dithering up to 8$^{\prime \prime}$ between each observation in a random pattern.

%\begin{figure*}

%\includegraphics[scale=0.7]{code/ks_plot1.png}
%\label{fig:fields}
%\caption{The Gemini GSAOI images of the three fields. Red circles indicate the unsaturated stars used to make the PSF profiles shown in Figure \ref{fig:psfs}}

%\end{figure*}

% \begin{table*}
% \caption{Gemini Observations}
% \begin{tabular}{lclcc}
% Field  &Center RA, Dec &  Dates & Total integration\\ 
%   Name                  &                          &   observed$^{*}$ & time (min)\\\hline\hline
% ES1-C      &00:35:16.8, -44:01:25 &   2014-12-03         &72.5 \\
% XMM-C2  & 02:27:42.0, -04:33:51&   2014-12-10,11      &32.5\\
% CDFS-C   & 03:29:13.8, -28:03:15&   2015-01-04         &65.0\\\hline
% \end{tabular}
% \label{tab:gemsobs}

% \noindent
% $^{*}$ a few observations were also obtained on other dates, but could not be added into the final mosaic 
% due to poor sky determination.
% \end{table*}

%%%%%%%%%%%%%%%%%%%%%%%%%%%%%%%%%%%%%%%%%%%%%%%
\begin{deluxetable}{ccccc}[t!]
\tablecaption{Gemini Observations Summary  \label{tab:gemsobs}}
\tablecolumns{5}
\tablewidth{0pt}
\tablehead{
\colhead{Field} & \colhead{RA} & \colhead{Dec.} & \colhead{Dates} & \colhead{Time} \\
\colhead{} & \colhead{(J2000)} & \colhead{(J2000)} & \colhead{} & \colhead{(minutes)}
}
\startdata
ES1-C   & 00:35:16.8 & -44:01:25 & 2014 Dec 03 & 72.5 \\
XMM-C2  & 02:27:42.0 & -04:33:51 & 2014 Dec 10-11 & 32.5\\
CDFS-C  & 03:29:13.8 & -28:03:15 & 2015 Jan 04 & 65.0\\
\enddata
\tablecomments{A few observations were also obtained on other dates but could not be added into the final mosaic due to poor sky determination.}
\end{deluxetable}
%%%%%%%%%%%%%%%%%%%%%%%%%%%%%%%%%%%%%%%%%%%%%%%%

% \begin{table*}
% \caption{Adaptive optics parameters and PSF properties of the observations.}
% \begin{tabular}{lccccc}
% Field Name &LGS Strehl& $r_0^{\ddag}$& PSF FWHM & PSF model & PSF model\\ 
%                     &($K$-band)$^{\dag}$ & (cm) & (mas) & weights$^{\mathsection}$& FWHM (pixels)$^{\mathparagraph}$ \\\hline\hline
% ES1-C      &14\%&18&94&1.88,~4.58,~15.2&0.14,~0.54,~0.32 \\
% XMM-C2  & 0.5\%&10 &155&1.88,~4.58,~15.2&0.14,~0.54,~0.32$^{*}$ \\
% CDFS-C   &12\% &13&70&1.73,~3.96,~10.0,~22.26&0.22,~0.34,~0.19,~0.24\\\hline
% \end{tabular}
% \label{tab:aoobs}
% \noindent
% $^{\dag}$ the mean of the $H$-band laser guide star Strehl ratio {\bf (the ratio of the peak of the theoretical PSF to that measured)} obtained from the image headers multiplied by the ratio of wavelengths to the power 1.2 to obtain an estimate of the $K$-band Strehl {\bf ratio}.
% \noindent
% $^{\ddag}$ the mean of the characteristic length scale of atmospheric turbulence measured at 500~nm (also from the image headers, {\bf obtained by analyzing the power spectrum of the corrections sent to the deformable mirror)}.
% \noindent
% $^{\mathsection}$ FWHM of the circular Gaussians in pixels that are coadded to form the PSF model after the weights are applied.
% \noindent
% $^{\mathparagraph}$ weights of the circular Gaussians that are coadded to form the PSF model. The weights are normalized to unity.
% \noindent
% $^{*}$ the PSF model from the ES1-C field was used for XMM-C2 due to the relatively poor quality of the observations in that field.
% \end{table*}

\subsection{Radio observations}

The ES1-C field was observed in four runs with the Australia Telescope Compact Array (ATCA) at 8.4~GHz, 2016 March 10-13. We obtained 56~hr of data. 
The XMM-C2 field was observed in a single pointing with the VLA in A-configuration as part of project VLA/14A-353 over the frequency range 8-12~GHz. Twelve executions of a single 1~hr scheduling block were observed in filler time, each spending 12.7~min on the XMM-C2 field, for a total time on source of 152~min.

\subsection{Optical observations in ES1}\label{sec:servs_i}

The ES1 field of SERVS was observed with the Cerro Tololo Inter-American Observatory (CTIO) Mosaic II camera in $i$-band in two runs, one on 2009 Nov 25, and the other from 2010 Sep 30 to 2010 Oct 4. Twelve pointings were observed so as to cover the whole SERVS ES1 field  (with the ES1-C field observed in the 2010 observations). 

\section{Data reduction and analysis}

\subsection{Gemini image reduction}

The images were run through the standard Gemini GSAOI software steps. Dark subtraction and flat fielding were
applied using appropriate observatory calibration files found in the Gemini archive.
The FITS files were then split up into individual
detector arrays. The data were then used to form a median background image, which was then subtracted, and the images were then shifted (using a bright, but unsaturated object as a reference) and averaged together, with the highest and lowest values for each pixel being rejected from the stack.
Array 2 was found to suffer from the
reset anomaly described in Schirmer et al.\ (2015). A similar solution was applied whereby the
median dark from the other images in the stack was scaled before being subtracted from the first frame. 

Sources in the reduced image for each array were then
matched to detections in the VIDEO survey (Jarvis et al.\ 2013).
The 
%plate 
pixel scale of the image given in the header (0.0194 arcsec/pixel)  was checked and was found to be correct within the uncertainty of the VIDEO positions, however, small corrections were applied to the rotation
and center of each array to align them with the VIDEO reference frame, using about ten objects per detector to perform the fitting. Within the scatter in our reference astrometry of $\approx 0\farcs05$, no evidence of non-linear distortions was found. 

\begin{figure*}
\includegraphics[clip=true, trim=0.95cm 0.9cm 4.9cm 2.5cm, width=\textwidth]{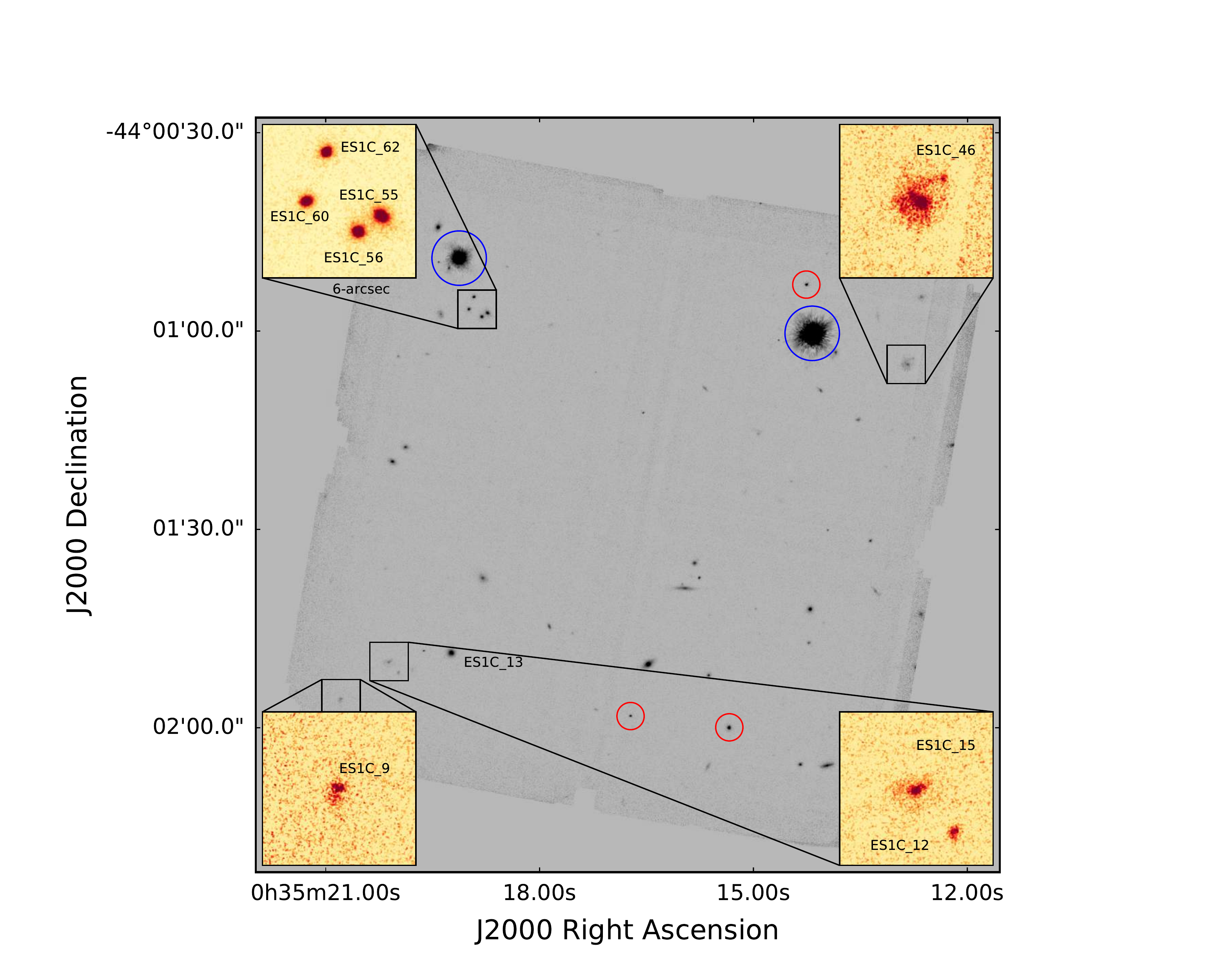}
\caption{The GSAOI image of the ES1-C field. Objects of interest due to their detection at other wavebands discussed in Section \ref{sec:herschel}, or with noteworthy morphologies (\ref{sec:other}) are shown as insets, each measuring 6$^{\prime \prime}$ per side.  The red circles indicate the stars used to determine the PSF in the field, and the blue circles show those used as natural guide stars for the adaptive optics system (one is off the image). Note that the guide stars were saturated, so they cannot be used for PSF determination.\\}
\label{fig:es1c}
\end{figure*}

\begin{figure*}
\includegraphics[clip=true, trim=0.95cm 0.9cm 4.9cm 2.5cm, width=\textwidth]{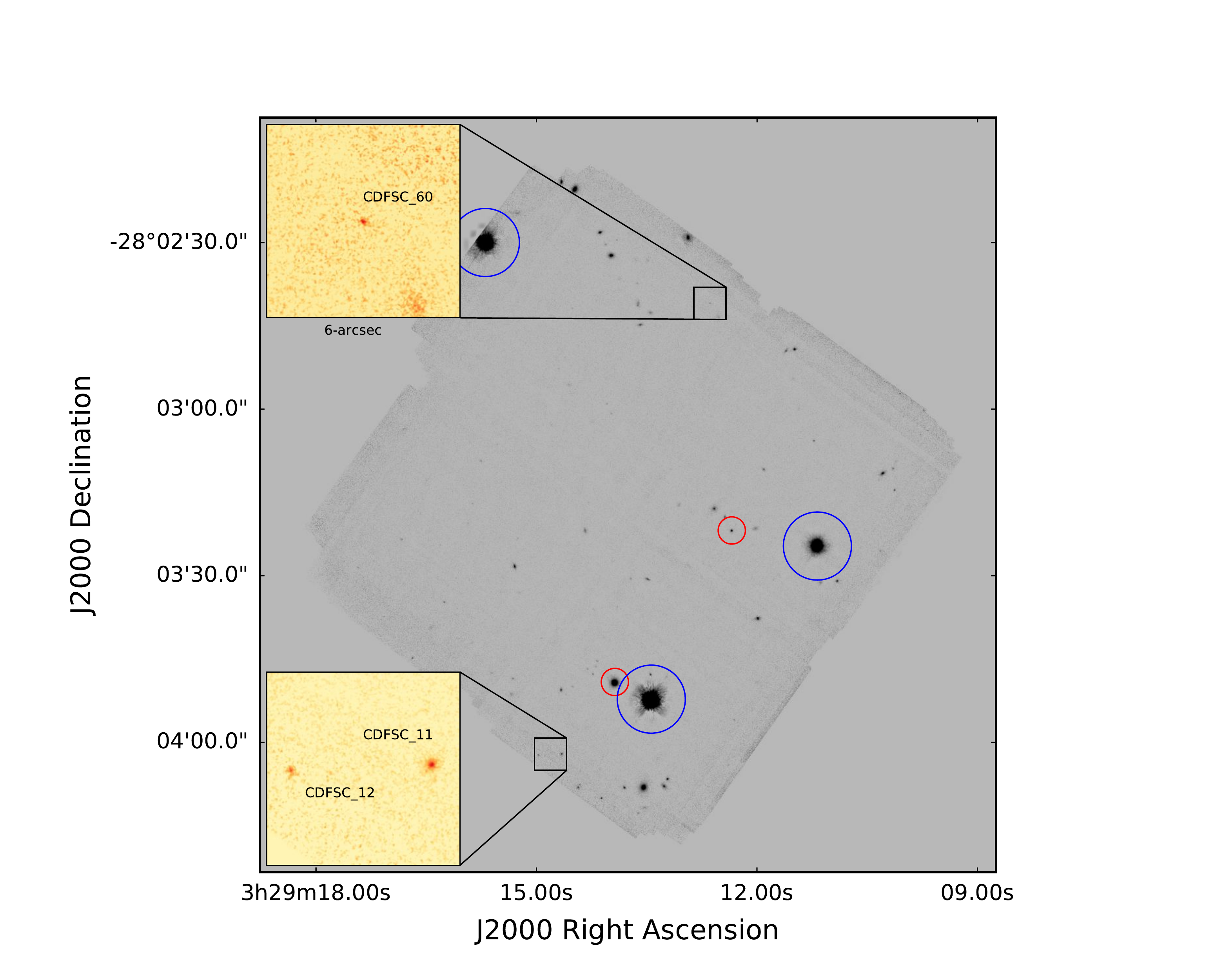}
\caption{The GSAOI image of the CDFS-C field. Three $z\sim 3$ compact galaxies (\ref{sec:other}) are shown in the insets, each measuring 6$^{\prime \prime}$ per side. The red circles indicate the stars used to determine the PSF in the field, and the blue circles show those used as natural guide stars for the adaptive optics system. Note that the guide stars were saturated, so they cannot be used for PSF determination.\\}
\label{fig:cdfsc}
\end{figure*}

\begin{figure*}
\includegraphics[clip=true, trim=0.95cm 0.9cm 4.9cm 2.5cm, width=\textwidth]{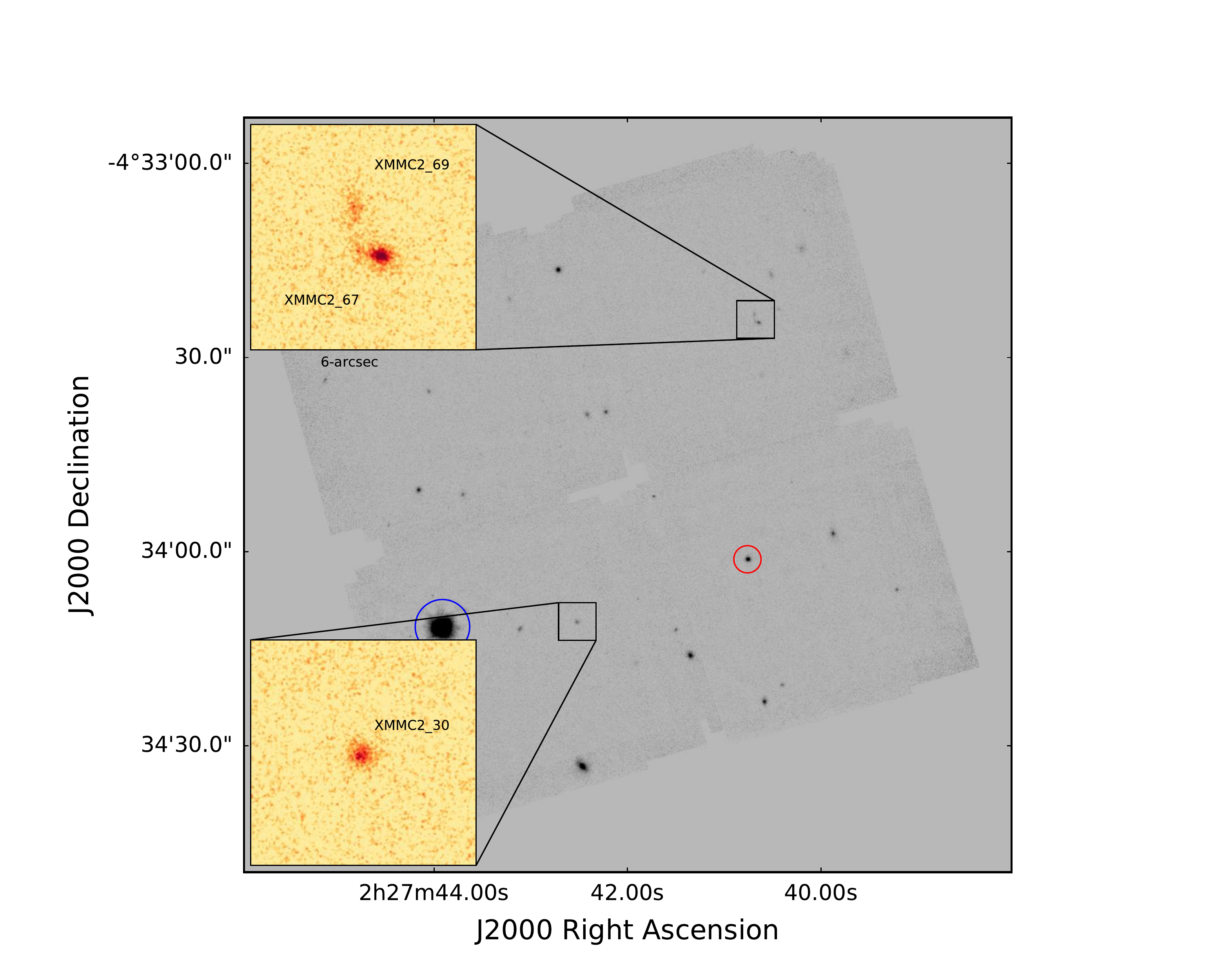}
\caption{The GSAOI image of the XMM-C2 field. Objects of interest due to their detection at other wavebands discussed in Section \ref{sec:herschel} are shown as insets, each measuring 6$^{\prime \prime}$ per side. The red circle indicates the star used to determine the PSF in the field, and the blue circle identifies one of the stars used as a natural guide star for the adaptive optics system (the other two are off the image). Note that the guide star was saturated, so it cannot be used for PSF determination.\\}
\label{fig:xmmc2}
\end{figure*}

As a check, we also fit the ES1-C image (which has the highest source density) for distortions using {\sc scamp} (Bertin 2006) with the VIDEO positions as a reference and confirmed that any remaining distortion is $<2$\% in linear scale across the frame. This results in a largest shift of 0.5$^{\prime \prime}$, consistent with that found by Schirmer et al.\ (2015).  As we are not using the data for high precision 
astrometry or photometric applications, we thus decided not to correct the frames for distortion. The Montage software (Berriman et al.\ 2002) was then used to combine the four arrays back into a single image. Flux calibration was obtained by scaling to the VIDEO catalog for isolated objects. The observations in the 
deepest two fields (ES1-C and CDFS-C) reach a 5$\sigma$ depth of mag$_{\rm AB} \approx 24.6$ for point sources. The XMM-C2 data, which were observed for a shorter time, reach a 5$\sigma$ depth of $\approx 24.0$. Grayscale images of the fields are shown in Figures \ref{fig:es1c}, \ref{fig:cdfsc}, and \ref{fig:xmmc2}.

%There remain some small problems in the overlap
%regions, notably one point source has a double appearance in the combined image, indicating an error of
%$\approx$2 pixels (0.04 arcsec) in the astrometry. 
%Objects in the overlap regions were therefore analysed on each chip separately. 

%@arxiver{ES1-C_galaxies_v4_compressed.pdf,es1c_h2.pdf,xmmc2_h1.pdf}

\begin{figure}
\centering
\includegraphics[clip=true, trim=1.95cm 4cm 1cm 3.5cm, width=3.5in]{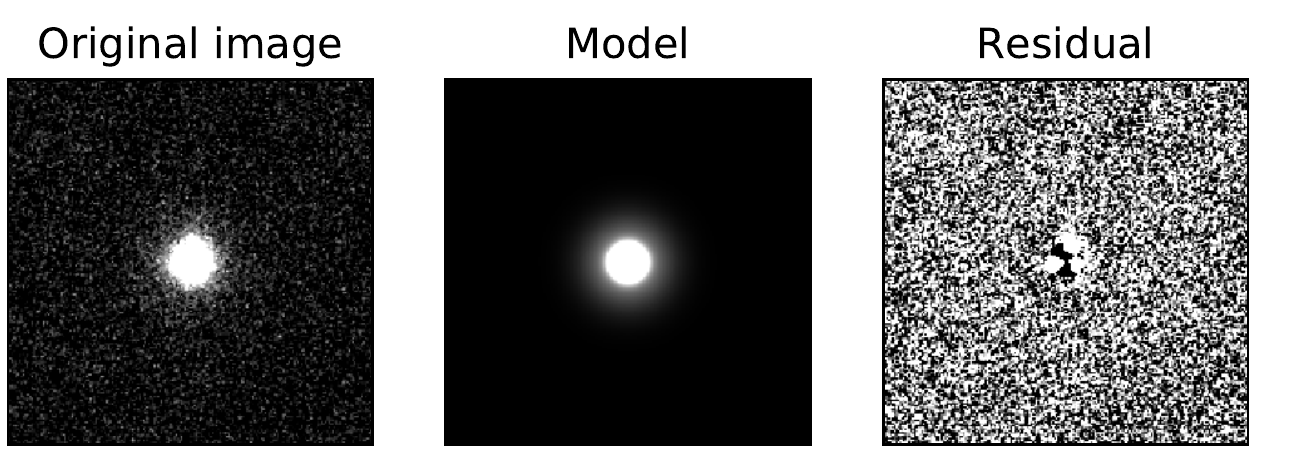}
\includegraphics[clip=true, trim=1.95cm 4cm 1cm 3cm, width=3.5in]{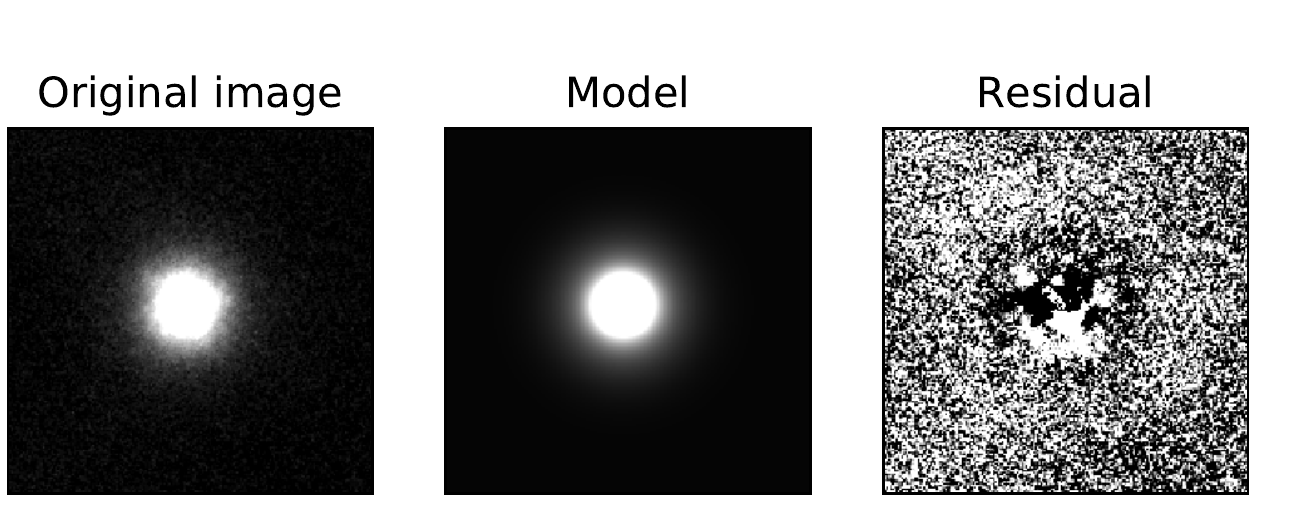}
\caption{The PSFs used in the Tractor fitting for ES1-C (top three images) and CDFS-C (bottom three images). From left to right, one of the PSF stars, the PSF model and the residual. The stars used to estimate the PSF are shown in Figures \ref{fig:es1c} and \ref{fig:cdfsc}. The cutouts are 4$^{\prime \prime}$ on a side, in the top figure the greyscale range is from 0 to 100 counts in the left two images, and -10 to 10 in the residual image, and in the bottom (where a brighter PSF star was available) from 0 to 500 and -10 to 10. Note that the data quality in the XMM-C2 field was low compared to that in the other two fields, so we used the ES1-C PSF for the analysis of those data.\\}
\label{fig:psfs}
\end{figure}

\subsection{Radio data processing}

Standard data reduction of the ATCA data was performed in MIRIAD (Sault, Teuben \& Wright 1995). Phase only self-calibration was applied, and images were made with natural and uniform weighting. In the naturally-weighted map we obtained an RMS noise of 1.8~$\mu$Jy~beam$^{-1}$ with a synthesized beam size of $2\farcs1 \times 1\farcs9$ at a position angle (PA) of 0$^{\circ}$. The uniformly-weighted map has an RMS noise of 2.3~$\mu$Jy~beam$^{-1}$ with a synthesized beam size of $1\farcs9 \times 1\farcs2$, also at PA = 0$^{\circ}$. 

The VLA data were reduced in CASA (McMullin et al.\ 2007) following standard procedures. No self calibration was applied. The synthesized beam size of our final image was $0\farcs21 \times 0\farcs16$ (naturally weighted) at PA = 48$^{\circ}$. The image reaches an RMS noise of 0.45~$\mu$Jy~beam$^{-1}$.

\subsection{Optical data processing}
Data from the 2010 CTIO run in ES1 were reduced at the Cambridge Astronomical Survey Unit; those from 2009 were reduced using the IRAF mscred package. The data were photometrically calibrated by cross-matching to DES (Dark Energy Survey Collaboration et al.\ 2016) data available from the National Optical Astronomical Observatory archive. The resulting image reaches an average $5\sigma$ depth of $i\approx 25$ mag.

\subsection{Point Spread Functions}
\label{sec:psfs}

We show the PSFs of stars in the ES1-C and CDFS-C fields in Figure \ref{fig:psfs} using the stars circled in red in Figures \ref{fig:es1c} and  \ref{fig:cdfsc}. The small number of unsaturated stars makes it difficult to assess the level of PSF variation across the field, though our measurements suggest it is small, $\sim 10$\% of the FWHM. We estimated the Strehl ratio for our observations by comparing the peak of a PSF model evaluated at the same pixel phase as our brightest PSF in each field to the observed peak value (in a 3\farcs8 $\times$ 3\farcs8 box). We used a simple obscured aperture model for the PSF:
\begin{equation}
I(\nu) = \frac{1}{(1-f^2)^2}\left[\frac{2 J_1(\nu)}{\nu}-f^2\frac{2J_1(f\nu)}{f\nu}\right]^2,
\end{equation}
where $\nu = \theta \pi D/ \lambda$, $\theta$ is the off-axis angle, $D=8.1$m is the effective diameter of the Gemini mirror, $f=0.123$ is the fractional radial obscuration of the aperture caused by the central hole (values of $D$ and $f$ are from Turri et al.\ 2018), and $J_1$ is the first order Bessel function.

In order to use the Tractor photometry code (Section \ref{sec:tractor}) we modeled the PSFs as sums of up to four circular Gaussians of different FWHM, modeling the stars in each field separately and averaging the results for each field. The Gaussians for a given star were all centered on a single position, which was allowed to vary during the fitting. The best fit was determined by minimizing the $\chi$-squared of the residuals.  The Strehl ratio, FWHM and the weights of the best-fitting Gaussian components used in each field are listed in Table \ref{tab:aoobs}. In ES1-C the PSFs were well fit by three circular Gausssians, and in CDFS-C four Gaussians were needed to capture the extended wings of the PSF. The XMM-C2 field was taken in significantly poorer conditions than the other two fields, and with less integration time, rendering an accurate PSF fitting not worthwhile, so we adopted the same PSF parameters as the ES1-C field for the purposes of the Tractor fitting.
  
\subsection{Optical to near-infrared photometry}\label{sec:tractor}

We used the Tractor (Hogg \& Lang 2013; Lang, Hogg \& Spergel 2016) to perform photometry matched between the high-resolution Gemini images and the lower resolution images from SERVS, VIDEO, CFHTLS, and other optical data (Tables \ref{tab:common} and \ref{tab:optical}).  The Tractor provides a convenient method of fitting model-based galaxy photometry without the need to re-sample images with widely differing pixel scales (Nyland et al.\ 2017). 
Sources in the Gemini data were fit either as point sources or, if resolved, with either exponential disk or de Vaucouleurs profiles (only a handful of the galaxies had high enough signal-to-noise ratios to reliably justify fitting a Sersic profile using the Tractor, so we did not perform fits to Sersic profiles). PSFs in all bands were modeled as mixtures of  circular Gaussians following the strategy described for the Gemini data in Section~\ref{sec:psfs}.

An initial catalog was compiled by running SExtractor (Bertin \& Arnouts 1999) on the Gemini data. This was then used to supply initial
guesses for the fit parameters in our Python implementation of the Tractor. 
A total of 165 objects were fit: 64 in ES1-C, 57 in CDFS-C, and 44 in the XMM-C2 field.
%The catalog was then looped through source-by-source
%and Tractor was used to fit 10$^{\prime \prime}$ cutout images centered on each source. 
% The decision to fit a point source instead of a galaxy model was made on the basis of the value of the SExtractor stellaricity parameter being $<0.1$, or the SExtractor size estimate being $<0\farcs05$.  
% The code was run twice, once fitting exponential disk models to resolved sources and once fitting de Vaucouleurs models, and the best fit (lowest reduced chi-squared) model was chosen for each object. Objects with radii consistent with zero from either run were refit with point source models. The code was then run a final time using the best models for each galaxy.
% Tractor convolves the source model with the PSFs at other wavelengths to obtain accurate multiband photometry. {\bf The final photometry is given in Table \ref{tab:optir_phot}.}
% More details of our implementation of the Tractor code may be found in Nyland et al. (2017).
We looped through our input catalog on a source-by-source basis,  extracting a 10$^{\prime \prime}$ cutout image at each source location. For each source in our input catalog, we fit either a point source or resolved surface brightness profile model.  
Point source models were fit for sources with a SExtractor stellaricity parameter $<0.1$ or SExtractor size estimate $<0\farcs05$.  Sources not meeting these criteria were modeled with either an exponential disk or de Vaucouleurs profile. All sources were fit for position and flux, and resolved sources for ellipticity and position angle\footnote{We note this differs from the procedure described in Nyland et al.\ (2017), where the the shape and position parameters were held fixed to improve speed and stability, and the Tractor was used only to fit the fluxes of the objects.}. 

The Tractor operates by convolving the source model with the PSF model at each band and then performing an optimization to determine the image flux, position, and shape properties.  
The code was run twice, once fitting exponential disk models to resolved sources and once fitting de Vaucouleurs models, and the best fit (lower reduced chi-squared) model was chosen for each object. Objects with radii consistent with zero from either run were refit with point source models. The final photometry is given in Table \ref{tab:optir_phot}.
Further details on our implementation of the Tractor image modeling code may be found in Nyland et al. (2017).

% \begin{table*}
% \caption{Multiwavelength data common to all three fields}
% \begin{tabular}{lcr}
% Bands &  Depth$^{\dag}$ & Survey \\\hline\hline
% $Z,Y^{*},J,H,K_s$ & 26.3,25.2,25.1,24.6,24.1 & VIDEO DR4 (Jarvis et al. 2013)\\
%  3.6$\mu$m,4.5$\mu$m  & 23.7,23.7&SERVS (Mauduit et al. 2012)\\
%  5.8$\mu$m, 8.0$\mu$m, 24$\mu$m, 70$\mu$m & 20.3,20.5,18.6,16.3&SWIRE (Lonsdale et al. 2003)\\
%  250$\mu$m, 350$\mu$m & 16.3,16.3& HerMES (Oliver et al. 2012)\\\hline
% \end{tabular}
% \label{tab:common}     

% \noindent
% $^{\dag}$Approx 3$\sigma$ depth (AB mags)

% \noindent
% $^{*} $The CDFS-C field did not have $Y$-band data available from VIDEO at the time of writing. 

%\end{table*}      

%%%%%%%%%%%%%%%%%%%%%%%%%%%%%%%%%%%%%%%%%%%%%%%
\begin{deluxetable*}{cccc}
\tablecaption{Multi-band Infrared Data Common to All Fields  \label{tab:common}}
\tablecolumns{4}
\tablewidth{0pt}
\tablehead{
\colhead{Bands} & \colhead{Depth$^{a}$} & \colhead{Survey$^{b}$} & \colhead{Ref.}
}
\startdata
$Z, Y^{a}, J, H, K_{\rm s}$ & 26.3, 25.2, 25.1, 24.6, 24.1 & VIDEO DR4 & Jarvis et al. 2013\\
 3.6$\mu$m, 4.5$\mu$m  & 23.7, 23.7 & SERVS & Mauduit et al. 2012 \\
 5.8$\mu$m, 8.0$\mu$m, 24$\mu$m, 70$\mu$m & 20.3, 20.5, 18.6, 16.3 & SWIRE & Lonsdale et al. 2003\\
 250$\mu$m, 350$\mu$m & 16.3, 16.3 & HerMES & Oliver et al. 2012\\
\enddata
\tablenotetext{a}{All depths are given at approximately the $3\sigma$ level (AB mag).}
\tablenotetext{b}{The CDFS-C field did not have $Y$-band data available from the VIDEO survey at the time of writing.}
\end{deluxetable*}
%%%%%%%%%%%%%%%%%%%%%%%%%%%%%%%%%%%%%%%%%%%%%%%%

% \begin{table*}
% \caption{Optical data}
%     \begin{tabular}{lcc}
%       Field &Optical Data Sources & Approx 3-$\sigma$ depths (AB)\\\hline\hline
%       ES1-C & DES $g$- and $r$-band; SERVS ancillary $i$-band (this paper)& $g\approx 25.0$, $r\approx 25.0$, $i\approx 25.6$\\
%       XMM-C2&CFHTLS D1;  $u,g,r,i,z$ &  $u\approx 26.4$,  $g\approx 26.1$,$r\approx 25.6$,$i\approx 25.3$,$z\approx 25.0^{*}$\\
%       CDFS-C &SWIRE ancillary optical &  $r\approx 25.0$, $g\approx 25.0$, $u\approx 25.5$\\ \hline
%      \end{tabular}
% \label{tab:optical}

%  \noindent
%  $^{*}$ 80\% completeness values from the  T0007 Final Release document \footnote{http://terapix.iap.fr/cplt/T0007/doc/T0007-doc.pdf}.
    
% \end{table*}   

%%%%%%%%%%%%%%%%%%%%%%%%%%%%%%%%%%%%%%%%%%%%%%%
\begin{deluxetable*}{ccc}
\tablecaption{Optical Data  \label{tab:optical}}
\tablecolumns{3}
\tablewidth{0pt}
\tablehead{
\colhead{Field} & \colhead{Optical Data Sources} & \colhead{Depth$^{a}$}
}
\startdata
ES1-C & DES $g$- and $r$-band; CTIO $i$-band (this paper)& $g\approx 25.0$, $r\approx 25.0$, $i\approx 25.6$\\
XMM-C2&CFHTLS D1;  $u,g,r,i,z$ &  $u\approx 26.4$,  $g\approx 26.1$,$r\approx 25.6$,$i\approx 25.3$,$z\approx 25.0$\\
CDFS-C &SWIRE ancillary optical &  $r\approx 25.0$, $g\approx 25.0$, $u\approx 25.5$\\
\enddata
\tablenotetext{a}{The CFHTLS depths are based on the 80\% completeness values from the  T0007 Final Release document (\protect{\href{http://terapix.iap.fr/cplt/T0007/doc/T0007-doc.pdf}{http://terapix.iap.fr/cplt/T0007/doc/T0007-doc.pdf}}). 
}
\end{deluxetable*}
%%%%%%%%%%%%%%%%%%%%%%%%%%%%%%%%%%%%%%%%%%%%%%%%

%%%%%%%%%%%%%%%%%%%%%%%%%%%%%%%%%%%%%%%%%%%%%%%
\begin{deluxetable*}{cccccc}[t!]
\tablecaption{Gemini Adaptive Optics and PSF Parameters  \label{tab:aoobs}}
\tablecolumns{5}
\tablewidth{0pt}
\tablehead{
\colhead{Field} & \colhead{Strehl Ratio} & \colhead{$\theta_{\rm FWHM}$} & \colhead{PSF Model FWHM} &\colhead{PSF Model Weights}\\
\colhead{} & \colhead{} & \colhead{(mas)} & \colhead{pixels} & \colhead{} \\
\colhead{(1)} & \colhead{(2)} & \colhead{(3)} & \colhead{(4)} & \colhead{(5)}}
\startdata
ES1-C  & 10\%  &  94  & 1.88, 4.58, 15.2        & 0.14, 0.54, 0.32 \\
XMM-C2 & 5\% &  155 & 1.88, 4.58, 15.2        & 0.14, 0.54, 0.32 \\
CDFS-C & 16\%  &  70  & 1.73, 3.96, 10.0, 22.26 & 0.22, 0.34, 0.19, 0.24\\
\enddata
\tablecomments{Column (1): Field name.  Column (2): The PSF Strehl ratio (the ratio of the peak of the measured PSF to the theoretical PSF), see text.  Column (3): FWHM of the observed PSF. Column (4): FWHM of the mixture of circular Gaussians used to form the PSF model after the weights are applied.  We note that the PSF model from the ES1-C field was used for XMM-C2 due to the relatively poor quality of the observations in that field. Column (5): Weights of each component of the mixture of circular Gaussians used to form the PSF model. The weights are normalized to unity. 
}
\end{deluxetable*}
%%%%%%%%%%%%%%%%%%%%%%%%%%%%%%%%%%%%%%%%%%%%%%%%

\begin{splitdeluxetable*}{ccccccccccccBcccccccccccBccccccccc}
\tablecaption{Optical through near-infrared Tractor photometry \label{tab:optir_phot}}
\tablehead{
\colhead{Source} & \colhead{RA} & \colhead{DEC} & \colhead{Radius} & \colhead{Ellip} & \colhead{PA} & \colhead{Model} & \colhead{$F_{CH2}$} & \colhead{$F_{CH1}$} & \colhead{$F_{Ks}$} & \colhead{$F_{H}$} & \colhead{$F_J$} & \colhead{$F_Y$} & \colhead{$F_{ZV}$} & \colhead{$F_{zO}$} & \colhead{$F_i$} & \colhead{$F_r$} & \colhead{$F_g$} & \colhead{$F_u$} & \colhead{$Ferr_{CH2}$} & \colhead{$Ferr_{CH1}$} & \colhead{$Ferr_{Ks}$} & \colhead{$Ferr_H$} & \colhead{$Ferr_J$} & \colhead{$Ferr_Y$} & \colhead{$Ferr_{ZV}$} & \colhead{$Ferr_{zO}$} & \colhead{$Ferr_i$} & \colhead{$Ferr_r$} & \colhead{$Ferr_g$} & \colhead{$Ferr_u$} & \colhead{$z_{m1}$}\\
\colhead{(1)} & \colhead{(2)} & \colhead{(3)} & \colhead{(4)} & \colhead{(5)} & \colhead{(6)} & \colhead{(7)} & \colhead{(8)} & \colhead{(9)} & \colhead{(10)} & \colhead{(11)} & \colhead{(12)} & \colhead{(13)} & \colhead{(14)} & \colhead{(15)} & \colhead{(16)} & \colhead{(17)} & \colhead{(18)} & \colhead{(19)} & \colhead{(20)} & \colhead{(21)} & \colhead{(22)} & \colhead{(23)} & \colhead{(24)} & \colhead{(25)} & \colhead{(26)} & \colhead{(27)} & \colhead{(28)} & \colhead{(29)} & \colhead{(30)} & \colhead{(31)} & \colhead{(32)}
}
\startdata
XMMC2\_1 & 36.92694 & -4.5758686 & 0.73 & 1.62& 46 & dev & 136.6 & 153.8 & 311.7 & 239.4 & 159.3 & 140.1 & 113.6& 96.7 & 74.4 & 43.2 & 12.0 & 3.6& 0.050 & 0.062 & 0.49 & 0.34 & 0.21 & 0.17& 0.086 & 0.039 & 0.014 & 0.010& 0.0080 & 0.014& 0.353 \\
XMMC2\_3 & 36.934834 & -4.574769 & 0.18& 1.30 & 269 & exp & 2.53& 2.85 & 5.79& 8.00 & 7.59& 8.81& 8.64& 6.14 & 6.03 & 5.01 & 3.13 & 1.21& 0.035& 0.044& 0.35& 0.24& 0.15 & 0.12& 0.061& 0.028 & 0.010 & 0.0074 & 0.0057 & 0.0096 & 0.021 \\
XMMC2\_5 & 36.919106 & -4.573089 & 0.20 & 1.23 & 8 & exp & 21.12 & 28.93 & 34.75& 25.17 & 15.10 & 12.40 & 9.29 & 7.62& 5.58 & 2.35 & 0.68 & 0.22 & 0.036 & 0.045& 0.35 & 0.24& 0.15 & 0.12& 0.061 & 0.028 & 0.010 & 0.0074 & 0.0057 & 0.0097 & 0.54 \\
\enddata
\tablecomments{Column (1): source name in the catalog. Columns (2) and (3): R.A. and Dec.. Column (4) the 
half-light radius of the fitted galaxy model in arcseconds (a value of zero indicates a point source was fit). Column (5) the axis ratio (major/minor) of the fitted model. Column (6) the position angle of the galaxy measured East of North. Column (7) the type of model fit (dev for a de Vaucouleurs profile, exp for an exponential disk, pt for a point source). Columns (8)-(19) flux densities in units of $\mu$Jy. $F_{Ks}$ , $F_{H}$ , $F_{J}$,  $F_{Y }$  and $F_{ZV}$ are derived from the VIDEO near-infrared data,  $F_{zO}$ ,  $F_{i}$ , $F_{r}$ , $F_{g}$ , $F_{u}$ are from optical data. $F_{CH1}$ and $F_{CH2}$ are the fluxes from the {\it Spitzer} IRAC 3.6 and 4.5$\,\mu$m bands, respectively. Columns (20)-(31) flux density errors in $\mu$Jy.  The flux errors are statistical errors from the surface brightness profile fits performed during the Tractor photometry and do not include calibration errors. Column (32) the best fit photometric redshift from EAZY. A full version of this table may be found in the online edition.}
\end{splitdeluxetable*}

\subsection{Mid- to far-infrared photometry}\label{sec:irphot}

There are several {\em Herschel} sources in the ES1-C and XMM-C2 fields detected in the HerMES survey (Oliver et al.\ 2012) by the SPIRE instrument at  250$\,\mu$m and 350$\mu$m (the CDFS-C field lacks {\em Herschel} detections). We used prior data from the SWIRE MIPS 24$\,\mu$m survey and our deep radio imaging to assess whether star formation or AGN emission dominates the infrared flux from each {\em Herschel} source (Table \ref{tab:observed}). We then used the Tractor to simultaneously fit point source models to the SWIRE 5.8, 8 and 24$\mu$m data and the HerMES 250 and 350$\mu$m data at the position of every optical/near-infrared galaxy with a mid-infrared or radio detection that lay within the contours of the {\em Herschel} source. The {\em Herschel} sources also appear in the HerMES DR4 catalogs (which use positional priors from {\em Spitzer} 24$\mu$m observations; Oliver et al.\ 2017, using a predecessor to the XID+ code of Hurley et al.\ 2017).  We have normalized the sums of the component Tractor fluxes to the sums of the components of the same {\it Herschel} detections in the HerMES DR4 catalog to ensure a consistent flux density scale. In some cases our attribution of fluxes amongst components has differed from that in the HerMES catalog; those cases are detailed in  Section \ref{sec:herschel}.

%%%%%%%%%%%%%%%%%%%%%%%%%%%%%%%%%%%%%%%%%%%%%%%%
\begin{figure*}[h!]
\centering
\includegraphics[clip=true, trim=0cm 1cm 0cm 2.75cm, width=\textwidth]{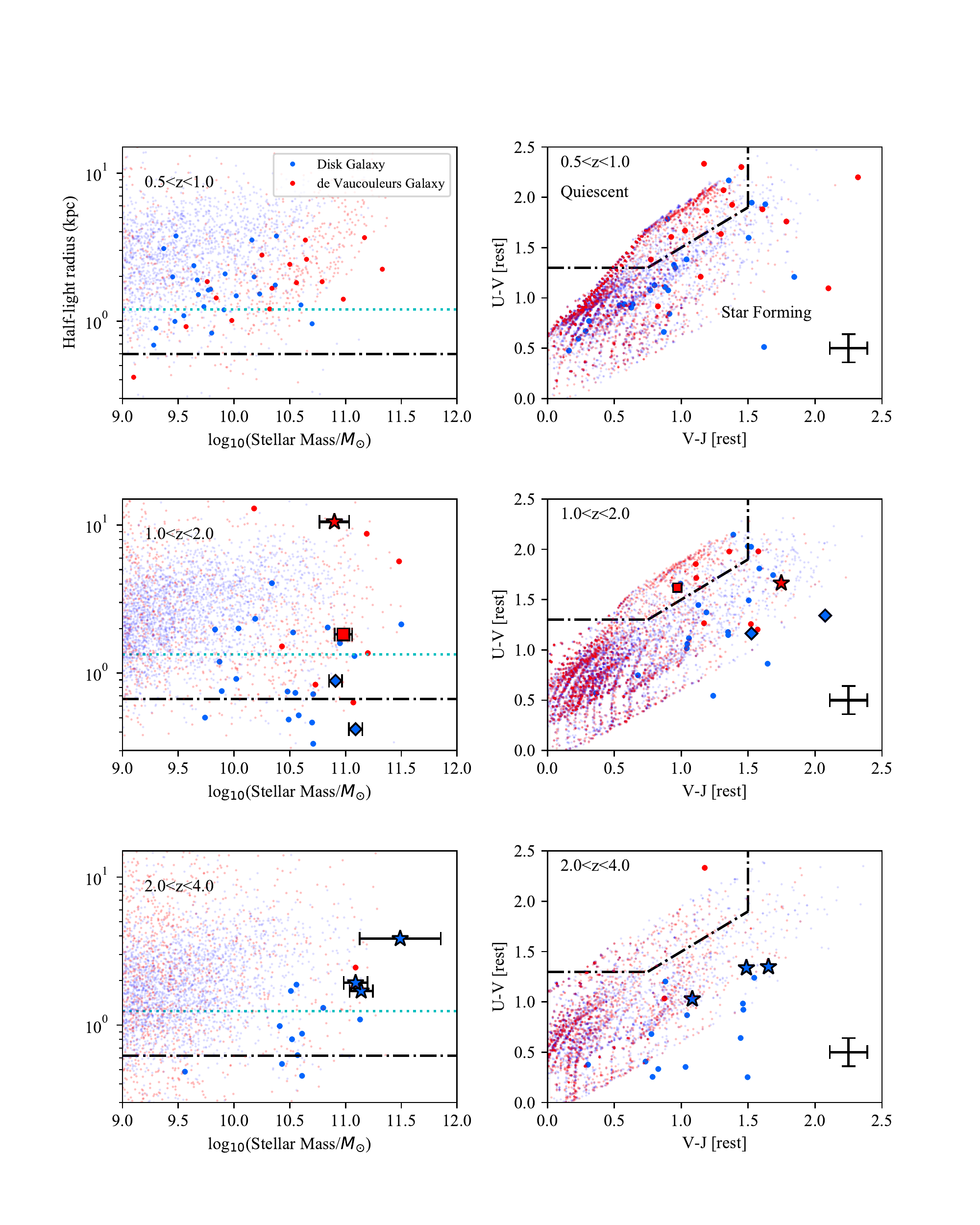}
\caption{Sizes, colors, and stellar masses of the galaxies identified in our study, split into three redshift bins. Galaxies best fit by disks are shown in blue; those best fit by de Vaucouleurs profiles are shown in red. Objects from the CANDELS COSMOS field (van der Wel et al.\ 2012; Nayyeri et al.\ 2017) are shown for comparison as faint dots, where we have colored objects fit with Sersic indices $<2.5$ in blue and those fit with Sersic indices $>2.5$ in red. Left column: half light radius vs stellar mass,  right column: rest-frame $U-V$ color vs.\ rest-frame $V-J$ color. 
Radio-quiet AGN are indicated with diamonds, radio-loud AGN with squares, and starbursts with stars. 
In the left column, the black dot-dashed line indicates the resolution of the Gemini images and the cyan dotted line the resolution of {\em HST}/WF3 in the F160W filter. In the right-hand column, the dot-dashed line indicates the divide between quiescent (upper left) and star-forming (lower right) galaxies, and the typical error bar is shown on the bottom right.}
\label{fig:sizes}
\end{figure*}
%%%%%%%%%%%%%%%%%%%%%%%%%%%%%%%%%%%%%%%%%%%%%%%%

\subsection{Photometric redshifts and stellar masses}
\label{sec:photoz}

Photometric redshifts were estimated using the EAZY software (Brammer, van Dokkum \& Coppi 2008). They were checked against photometric redshifts from the SERVS photometric redshift catalog (Pforr et al. in prep) for objects that are detected in the SERVS catalogs. In addition to the redshifts, the EAZY code was used to calculate the rest-frame $U-V$ and $V-J$ colors used in Figure \ref{fig:sizes}.

Stellar masses ($M_{*}$) were estimated by performing spectro-photometric fitting of stellar population models with the HyperZ code (Bolzonella et al.\ 2000) and using the ancillary scripts for $M_{*}$~calculation as in Daddi et al. (2005) and Maraston et al. (2006). As in these papers, data were fitted using a comprehensive suite of models (from Maraston~2005) for a variety of star formation histories, spanning from single bursts to constant star formation rates, each calculated for 221 ages from 0.001~Myr to 15000~Myr, and four metallicities below, at, and, above solar ([Z/H]$=$-1.35, -0.33, 0.0 and 0.35). Calculations were performed assuming no reddening, a Calzetti reddening law (Calzetti, Kinney \& Storchi-Bergman 1994), and a reddening law typical of the Small Magellanic Cloud (SMC). The no-reddening option is useful to check how much the derived stellar mass is affected by age-dust degeneracies, which may lead to and underestimate of  $M^{*}$~for star-forming galaxies (see Pforr, Maraston \& Tonini 2012). This is known as the ``iceberg effect'' (Maraston et al. 2010), when by fitting the youngest stellar populations with reddening, the estimated stellar mass is closer to the mass involved in the burst (the tip of the iceberg) rather than to the total one. The SMC law is included besides the Calzetti law as it is found to give good results in high-z, passive galaxies (Maraston et al. 2006; Kriek \& Conroy 2013). In the calculations, the EAZY photometric redshifts were assumed for constraining the fits.

\subsection{SED fitting}\label{sec:sedfit}
\begin{figure*}
\includegraphics[clip=true, trim=1.5cm 4cm 2cm 2.5cm, width=\textwidth]{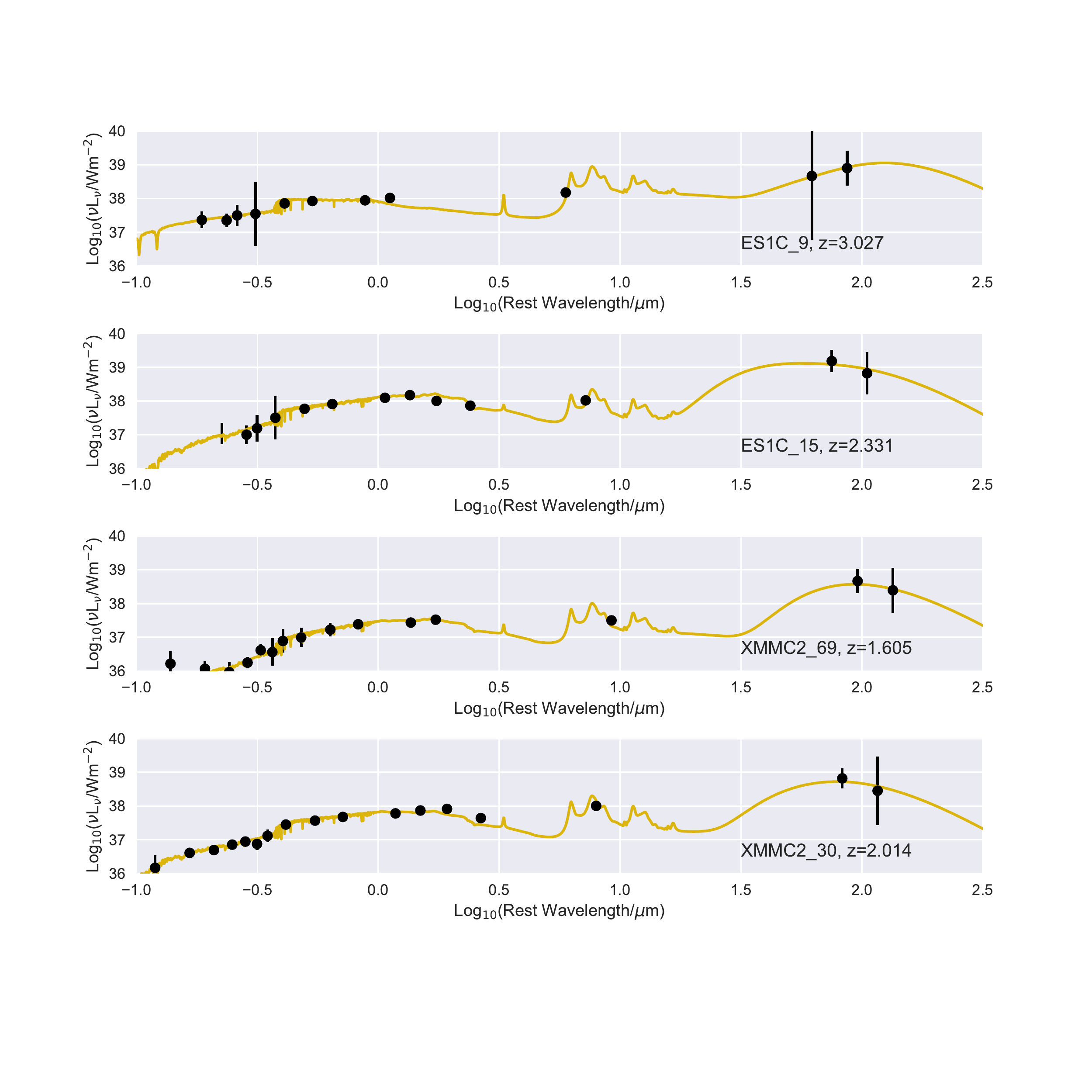}
\caption{SEDs of galaxies dominated by star formation. The gold solid lines are the fits from {\sc magphys}.
\\}
\label{fig:seds1}
\end{figure*}

We classified the sources detected in the {\em Herschel} images as pure starbursts or composite starbursts+AGN based on their mid-infrared colors (e.g.\ Lacy et al. 2004). In addition, we identify one radio-loud AGN that does not have significant far-infrared flux. The purely star forming galaxies were fit using the {\sc magphys} code (da Cunha, Charlot \& Elbaz 2008), which models the three emission components of the ISM (polycyclic aromatic hydrocarbons, warm dust and a cooler dust component) and UV emission from starlight in a self-consistent way (Figure \ref{fig:seds1}). The 
reduced $\chi^2$ values of the fits vary from 1.5 to 8.2.
All Herschel sources have infrared luminosities $L_{\rm IR}> 10^{12}L_{\odot}$ so qualify as Ultra-Luminous Infrared Galaxies (ULIRGs).
%(for ES1C\_9. 

\begin{figure*}
\includegraphics[clip=true, trim=1.5cm 2.4cm 2cm 2cm, width=\textwidth]{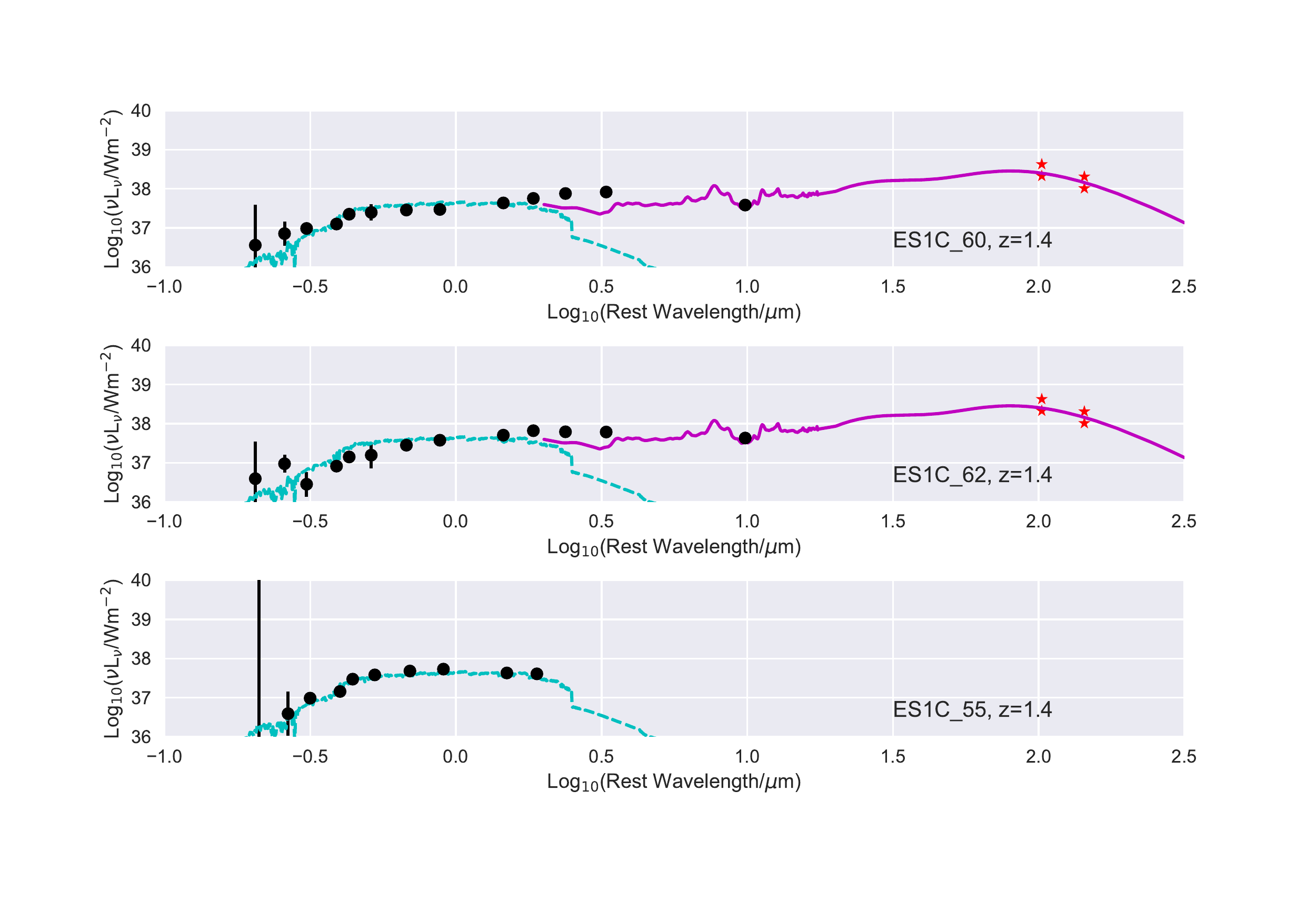}
\caption{SEDs of the candidate triple AGN in the ES1-C field. The cyan dashed lines are a single 1~Gyr stellar population from Maraston et al.\ (2005) and the solid magenta lines are the Composite-2 SED from Kirkpatrick et al.\ (2015). The Composite-2 SED was not fit to the radio galaxy (ES1C\_55) as it lacks detectable mid-infrared emission. The two sets of stars in the top two panels showing ES1C\_60 and ES1C\_62 indicate the total fluxes in the blended {\it Herschel} data (upper stars) and a 50\% attribution to each object (lower stars).\\}
\label{fig:seds2}
\end{figure*}

For relatively bright objects in the mid-infrared consistent with mixed AGN and star formation activity, we used the composite AGN/starburst templates from Kirkpatrick et al. (2015) to estimate the star formation rates, as shown in Figure \ref{fig:seds2}. 
We selected the closest template to the SEDs of our AGN, the ``Composite-2" SED (with an AGN contribution to the IR luminosity of 60\%). Although the mid-infrared excess from the AGN dust is larger in our objects than the template (suggesting a larger AGN contribution), the far-infrared SED matches well\footnote{We note that the Kirkpatrick et al.\ templates have too low of a far-infrared to mid-infrared flux ratio to fit our two most actively starbursting galaxies, probably because the sample used to make them was selected in the mid-infrared.}. We also provide estimates of star formation rates based on both the far-infrared emission and the radio emission using equations~4 and 6 from Bell (2003). The results are given in Table \ref{tab:derived} and details about each object are given in Section~\ref{sec:herschel}.

\subsection{The radio--far-infrared correlation}
% The radio -- far-infrared correlation (Helou et al.\ 1985) is a tight correlation between the radio continuum emission and the far-infrared emission in galaxies. First seen locally, it is now known to extend out to out least $\sim 2$ (e.g.\ Mao et al.\ 2011). 
The tight radio -- far-infrared correlation extends over three orders of magnitude in normal star-forming galaxies (Helou et al.\ 1985). First seen locally, it is now known to extend out to out least $z \sim 2$ (e.g.\ Ivison et al.\ 2010; Mao et al.\ 2011; Magnelli et al.\ 2015; Delhaize et al.\ 2017), though the nature of the redshift evolution of the relation remains a subject of debate.   
Here, we use the radio-infrared correlation as a diagnostic tool for identifying radio AGNs based on the presence of a significant excess of radio emission compared to the level expected purely from star formation. 

We use our new ATCA and VLA radio continuum data and the estimates of total infrared luminosity from our SED modeling ($S_{\rm IR}$) to estimate the logarithmic infrared to radio flux ratio, $q_{\rm IR}={\rm log_{10}}[(S_{\rm IR}/3.75\times 10^{12}\, {\rm W\,m^{-2}})/(S_{\rm 1.4\,GHz}/{\rm W\,Hz^{-1}\,m^{-2}})]$, following Ivison et al.\ (2010). $S_{\rm 1.4\,GHz}$ is calculated in the rest frame assuming a radio spectral index\footnote{We follow the convention for radio spectral index ($\alpha$) of $S \sim \nu^{\alpha}$, where $S$ is the flux, $\nu$ is the frequency.} of $\alpha = -0.8$.  We provide the $q_{\rm IR}$ values for our sources in Table~\ref{tab:observed}. The $q_{\rm IR}$ values indicate that most of the {\em Herschel} sources lie close to the typical value for star-forming galaxies of $q_{\rm IR}\approx 2.4$, certainly within the $\pm \approx 0.4$ dex scatter seen by Ivison et al.\ (2010) for {\em Herschel} sources. As expected, the  radio-quiet AGN pair ES1C\_60/ES1C\_62 lies above the relation ($q_{\rm IR}=3.1$), though when the infrared flux is reduced by 60\% to allow for the AGN contribution, it falls within the scatter {expected for normal star-forming galaxies}.

% \begin{table*}
% \caption{Profiles of galaxies in the quiescent and star forming regions of the UVJ plot \label{tab:bulgefrac}}
% \begin{tabular}{lccc}
%  Redshift Range & $0.5<z<1.0$ & $1.0<z<2.0$ & $2.0<z<4.0$\\\hline\hline
% Best fit by disk, star-forming   & 23/30 & 4/6 &  1/1\\
% Best fit by de Vaucouleurs, quiescent & 7/9&17/22 & 16/17\\\hline
% \end{tabular}
% \end{table*}

%%%%%%%%%%%%%%%%%%%%%%%%%%%%%%%%%%%%%%%%%%%%%%%
\begin{deluxetable}{ccc}
\tablecaption{Surface Brightness Profile Demographics  \label{tab:bulgefrac}}
\tablecolumns{3}
\tablewidth{0pt}
\tablehead{
\colhead{Redshift Range} & \colhead{Exponential} & \colhead{de Vaucouleurs}
}
\startdata
$0.5<z<1.0$ & 23/30 & 7/9\\
$1.0<z<2.0$ & 4/6   & 17/22\\
$2.0<z<4.0$ & 1/1   & 16/17\\
\enddata
\tablecomments{Redshift dependence of best-fitting surface brightness profile from our Tractor photometry.  Sources best fit by exponential profiles are disk galaxies with high star formation rates, while sources best fit by de Vaucouleurs profiles tend to be bulge-dominated, quiescent galaxies.}
\end{deluxetable}
%%%%%%%%%%%%%%%%%%%%%%%%%%%%%%%%%%%%%%%%%%%%%%%%

% %%%%%%%%%%%%%%%%%%%%%%%%%%%%%%%%%%%%%%%%%%%%%%%%
% \begin{figure*}
% \centering
% %\includegraphics[scale=0.8]{rad_mass.pdf}
% \includegraphics[clip=true, trim=0cm 1cm 0cm 2.75cm, width=\textwidth]{rad_mass.pdf}
% %\includegraphics[clip=true, trim=0.95cm 0.9cm 4.9cm 2.5cm, width=\textwidth]{CDFS-C_galaxies_v4_reduce.pdf}
% \caption{{\bf Sizes, colors and stellar masses of the galaxies in the fields, split into three redshift bins. Galaxies best fit by disks are shown in blue, those best fit by de Vaucouleurs profiles in red. Objects from the CANDELS COSMOS field (van der Wel et al.\ 2012; Nayyeri et al.\ 2017) are shown for comparison as faint dots, where we have colored objects fit with Sersic indices $<2.5$ in blue and those fit with Sersic indices $>2.5$ in red. Left column: half light radius vs stellar mass,  right column: rest-frame $U-V$ color vs rest-frame $V-J$ color. 
% Radio-quiet AGN are indicated with diamonds, radio-loud AGN with squares, and starbursts with stars. 
% In the left column,} the black dot-dashed line indicates the resolution of the Gemini images and the cyan dotted line the resolution of {\em HST}/WF3 in the F160W filter. In the right-hand column, the dot-dashed line indicates the divide between quiescent (upper left) and star forming (lower right) galaxies, and the typical error bar is shown on the bottom right.}
% \label{fig:sizes}
% \end{figure*}
% %%%%%%%%%%%%%%%%%%%%%%%%%%%%%%%%%%%%%%%%%%%%%%%%

\section{Results}

\subsection{Galaxies in the field}

Figure \ref{fig:sizes} shows the sizes, colors and stellar masses of the galaxies 
%in our three fields 
considered in our study for three different redshift ranges. The redshift bins were picked to roughly divide the study into epochs of later galaxy evolution ($0.5<z<1.0$ - the lower limit was picked as there are very few $z<0.5$ galaxies with masses in the same range as the higher-$z$ objects), peak galaxy formation ($1.0<z<2.0$) and early galaxy formation ($2<z<4$). (There are too few $z>4$ objects with reliable photo-zs to explore higher in redshift.)
%Measurements of the evolution of galaxy size as a function of galaxy type and mass are essential for constraining models of galaxy formation, and provide an important use case for MCAO (e.g.\  Sweet et al.\ 2017). 
Galaxy size evolution is closely tied to merger activity, as dry mergers can ``puff up" small galaxies (e.g.\ Bluch et al.\ 2012; Newman et al.\ 2012), and estimates of the galaxy size distribution as a function of redshift are thus important for our understanding of the importance of mergers in galaxy evolution. As seen in prior studies with {\em HST} (e.g. van der Wel et al.\ 2014, Shibuya et al.\ 2015, Allen et al.\ 2017), there is strong evidence for size evolution at a given stellar mass, particularly for galaxies with masses $<10^{11}$~M$_{\odot}$. The right-hand column of Figure \ref{fig:sizes}  shows the $U-V$ vs $V-J$ rest-frame color-color diagram that Whitaker et al.\ (2011) use to separate star-forming galaxies from quiescent galaxies. This figure shows that our morphological classifications are broadly in line with the expectations from this plot, namely that the quiescent galaxies include a large fraction with de Vaucouleurs profiles at all redshifts, whereas the star-forming galaxies are predominantly disks (Table \ref{tab:bulgefrac}). 

Comparing to the CANDELS results of van der Wel et al., shown as the faint dots in Figure \ref{fig:sizes}, we do see a larger fraction of compact star-forming galaxies, particularly at $z>2$. Only one of the 13 $z>2$ star-forming galaxies with masses $>10^{10}\,M_{\odot}$ in our study has a radius $>2$~kpc, compared to 337 out of 774 (43\%) in CANDELS-COSMOS. This may be a selection effect due to lower surface brightness sensitivity in the Gemini data causing us to miss extended disks, though at high redshift the radii of the galaxies are close to the {\em HST} resolution limit of $\approx 0.7$~kpc (HWHM), suggesting that the limited resolution of {\em HST} may also play a role. Deeper Gemini observations would help us to better understand this issue.

\subsection{Galaxies detected by Herschel/SPIRE}\label{sec:herschel}

The four {\em Herschel} sources (two in ES1-C and two in XMM-C2) are comprised of multiple galaxies, some with similar photometric redshifts and some with very different ones. The {\em Herschel} images suffer from source confusion, so blending of sources that may or may not be physically associated is common. Details of the {\em Herschel}  detections are given in Table \ref{tab:observed}; see Section \ref{sec:irphot} for details of the deblending. The SEDs were fit as described above in Section~\ref{sec:sedfit}.

%The purely star forming galaxies were fit using the {\sc magphys} code (da Cunha, Charlot \& Elbaz 2008), which models the three emission components of the ISM (polycyclic aromatic hydrocarbons, warm dust and a cooler dust component) and UV emission from starlight in a self-consistent way (Figure \ref{fig:seds1}). For objects relatively bright in the mid-infrared, indicating mixed AGN and star formation, we used the templates from Kirkpatrick et al. (2015) to estimate the star formation rates, as shown on Figure \ref{fig:seds2}). Specifically the ``Composite-2" SED (with an AGN contribution to the IR luminosity of 60\%) is the best fit. (Note that the Kirkpatrick et al.\ templates have too low of a far-infrared to mid-infrared flux ratio to fit our two most actively starbursting galaxies.) We use the estimates of total infrared luminosity from the fits to estimate the logarithmic infrared to radio flux ratio $q_{\rm IR}={\rm log_{10}}[(S_{\rm IR}/3.75\times 10^{12} {\rm Wm^{-2}})/(S_{\rm 1.4GHz}/{\rm WHz^{-1}m^{-2}})]$ (following Ivison et al.\ 2010), where $S_{\rm 1.4GHz}$ is calculated in the rest frame assuming a radio spectral index of -0.8, and show this in the last column of Table \ref{tab:observed}.
%Finally, we provide estimates of star formation rates based on both the far-infrared emission and the radio emission using the formulae of Bell (2003) (equations 4 and 6). The results are given in Table \ref{tab:derived} and details of of each object are given in Section\ref{sec:herschel}

\subsubsection{ES1C\_H1}

\begin{figure*}
\centering
\includegraphics[clip=true, trim=0cm 0cm 3cm 2cm, width=\textwidth]{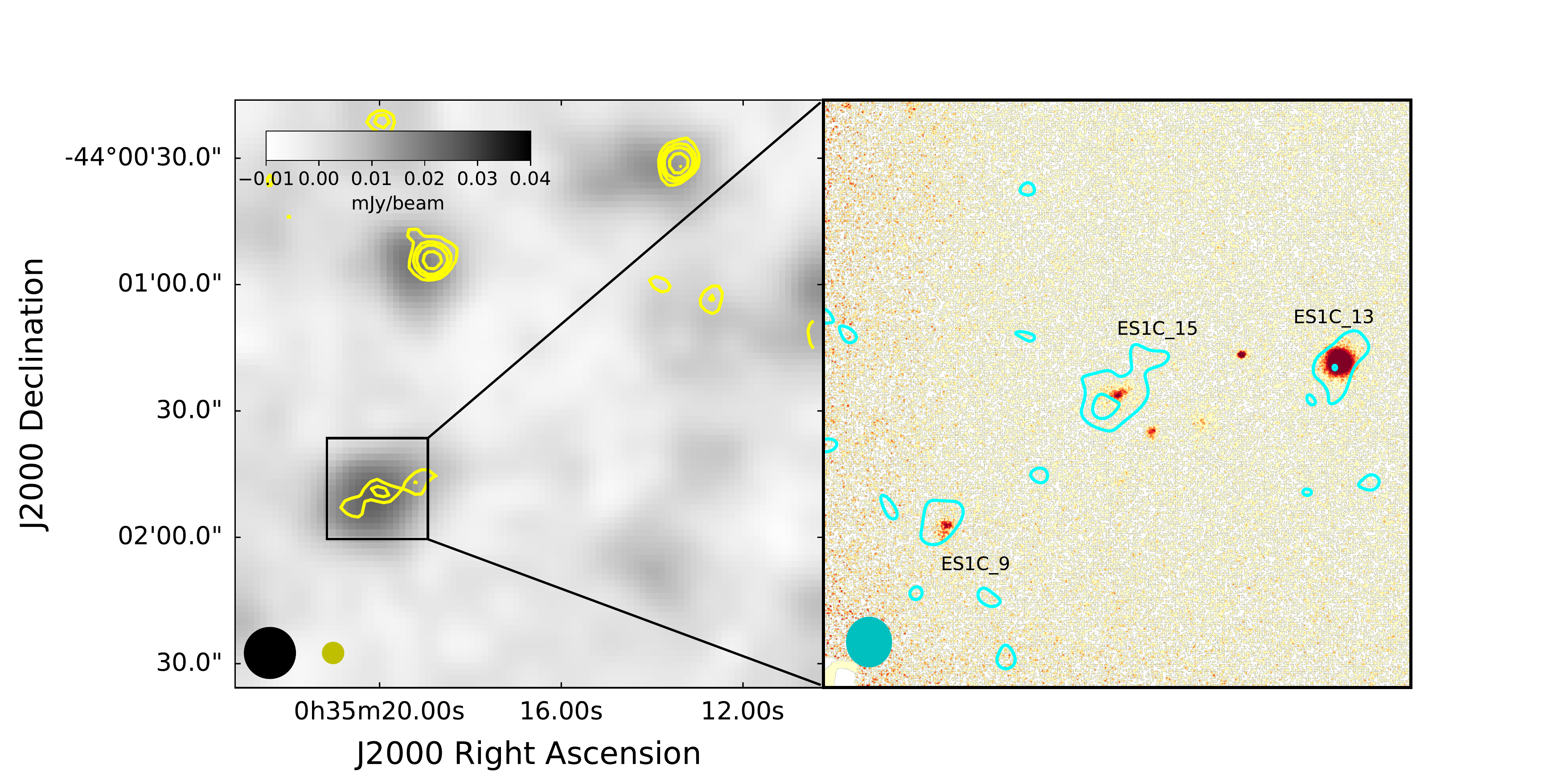}
\caption{ES1C\_H1. The left-hand panel shows a grayscale image of the Herschel/SPIRE 250$\mu$m image of the ES1-C field with overlaid contours  of the SWIRE MIPS 24$\mu$m in yellow (levels 0.1, 0.15, 0.2, 0.3 and 0.4~MJy~sr$^{-1}$). The right-hand panel shows a zoom in of the region around ES1C\_H1, showing the Gemini image overlaid ATCA 8.4~GHz contours from the naturally-weighted image in cyan at 3.6 and 7.2~$\mu$Jy~beam$^{-1}$. Beam sizes (FWHM) are indicated in the bottom-left of each image.}
\label{fig:ES1C-H1}
\end{figure*}

This source (Figure \ref{fig:ES1C-H1}) has three 24$\mu$m sources apparently associated with it, which are in turn associated with galaxies ES1C\_9, ES1C\_13, and ES1C\_15. ES1C\_9 ($z_{\rm phot}=3.0$) appears to have an irregular morphology, with a compact core and more diffuse emission to its south (Figure \ref{fig:es1c}, bottom left inset). It is brighter at 350$\,\mu$m than 250$\,\mu$m (Figure \ref{fig:seds1}), and barely detected at 10~GHz, consistent with its relatively high redshift. ES1C\_13 
($z_{\rm phot}=0.3$) is relatively bright at 24$\,\mu$m and detected at 10~GHz, but seems to not contribute to the {\it Herschel} source, consistent with its low redshift. ES1C\_15 ($z_{\rm phot}=2.3$) is another disturbed looking galaxy, possibly in a group (though ES1C\_12, the most nearby candidate companion, has a lower photometric redshift of 1.6). ES1C\_15 dominates the 250$\,\mu$m flux and it is detected at 10~GHz (Figure~\ref{fig:es1c}, bottom right and Figure \ref{fig:ES1C-H1}).  The peaks of both the infrared and radio emission are offset to the South-East of the stellar light by about $0\farcs5$. In the HerMES catalog, one source is between ES1C\_13 and ES1C\_15 (4HERMES~S250~SF~J003519.2-440145) and one between ES1C\_9 and ES1C\_15 (4HERMES~S250~SF~J003520.4-440151). %As we discussed above, both the Tractor fitting and the redshift of the galaxies suggest that the dominant contributors to the {\em Herschel} source are ES1C\_9 and ES1C\_15.

\begin{figure*}
\centering
\includegraphics[clip=true, trim=0cm 0cm 3cm 2cm, width=\textwidth]{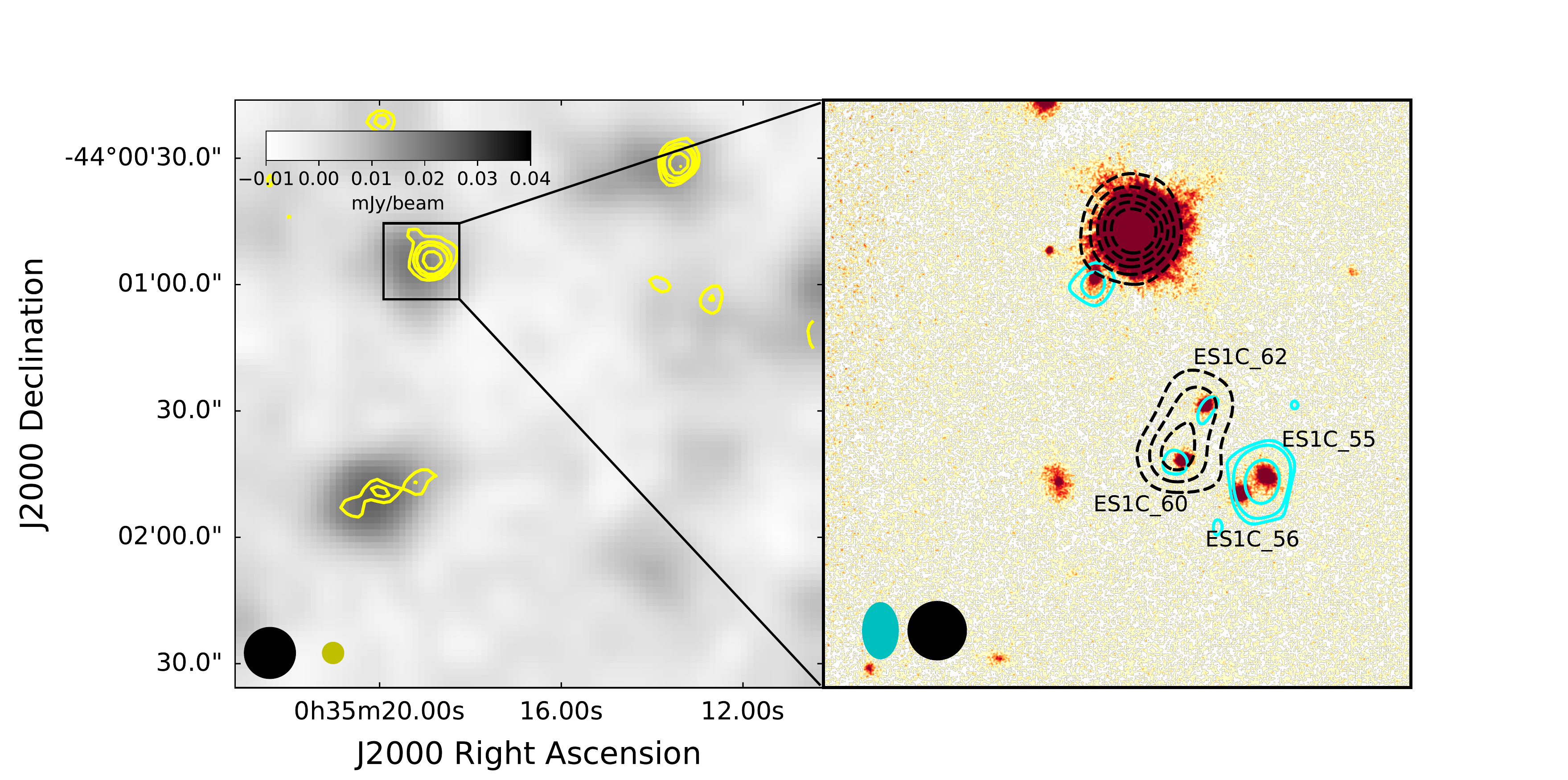}
\caption{ES1C\_H2. The left-hand panel shows a grayscale image of the Herschel/SPIRE 250$\,\mu$m image of the ES1-C field with overlaid contours  of the SWIRE MIPS 24$\mu$m in yellow (levels 0.1, 0.15, 0.2, 0.3 and 0.4~MJy~sr$^{-1}$). The right-hand panel shows a zoom in of the region around ES1C\_H2, showing the Gemini image, overlaid with ATCA 8.4GHz contours from the uniform-weighted image in cyan at 6, 12 and 60$\mu$Jy/beam , and SWIRE 8$\mu$m emission (dashed) in black (contour levels 7.75, 7.8, 7.85, 8.2 and 8.4~MJy~sr$^{-1}$). Beam sizes (FWHM) are indicated in the bottom-left of each image.}
\label{fig:ES1C-H2}
\end{figure*}

\begin{figure*}
\includegraphics[clip=true, trim=0cm 0cm 3cm 2cm, width=\textwidth]{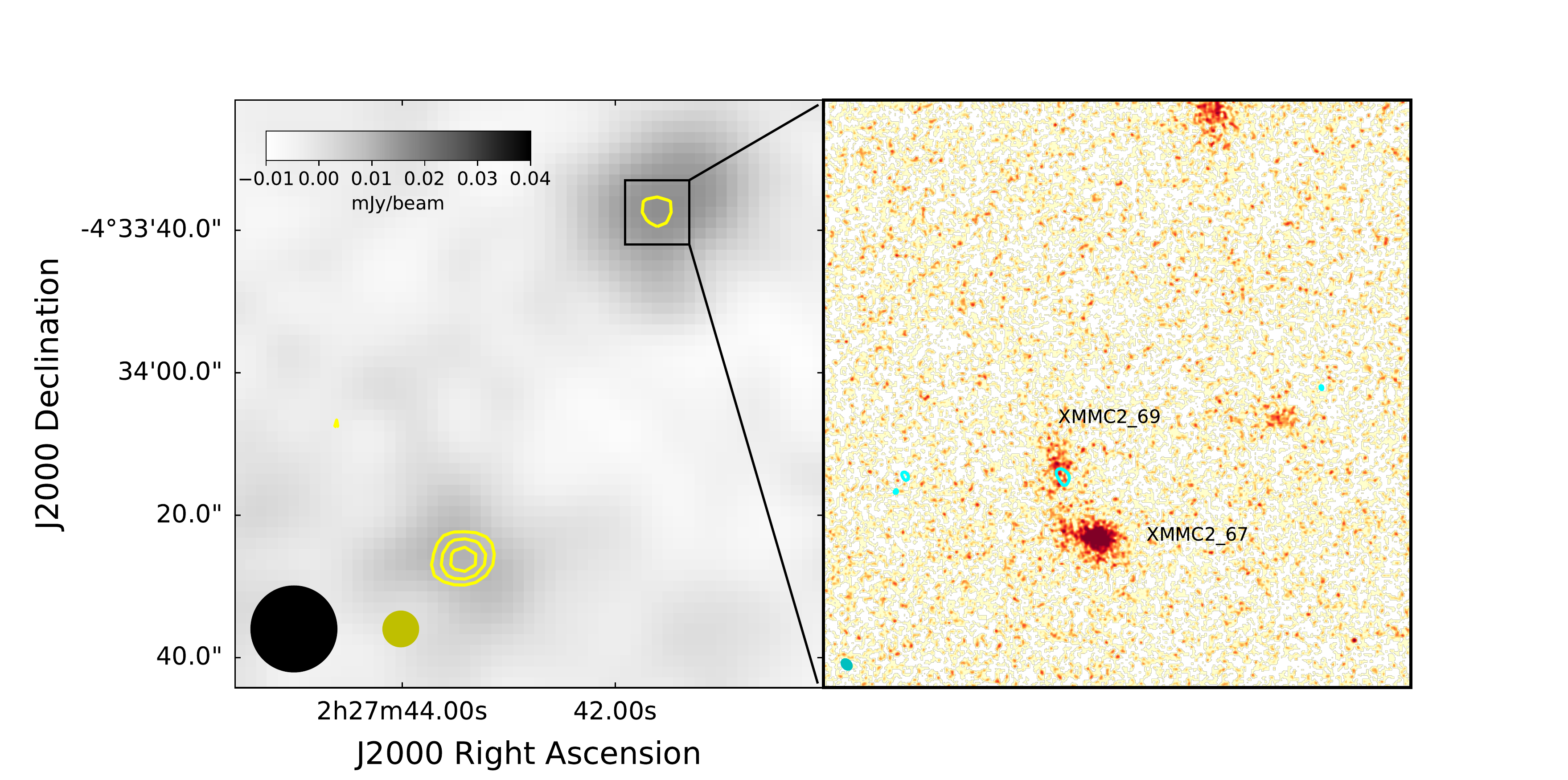}
\caption{XMMC2\_H1.  The left-hand panel shows a grayscale image of the Herschel/SPIRE 250$\mu$m image of the XMMC-2 field with overlaid contours  of the SWIRE MIPS 24$\mu$m in yellow (levels 0.1, 0.15 and 0.2~MJy~sr$^{-1}$) above background). The right-hand panel shows a zoom in of the region around XMMC2\_H1, showing the Gemini image overlaid with the 10$\,\mu$Jy~beam$^{-1}$ contour from the naturally-weighted VLA 10~GHz image in cyan. Beam sizes (FWHM) are indicated in the bottom-left of each image.}
\label{fig:XMMC2-H1}
\end{figure*}

\begin{figure*}
\includegraphics[clip=true, trim=0cm 0cm 3cm 2cm, width=\textwidth]{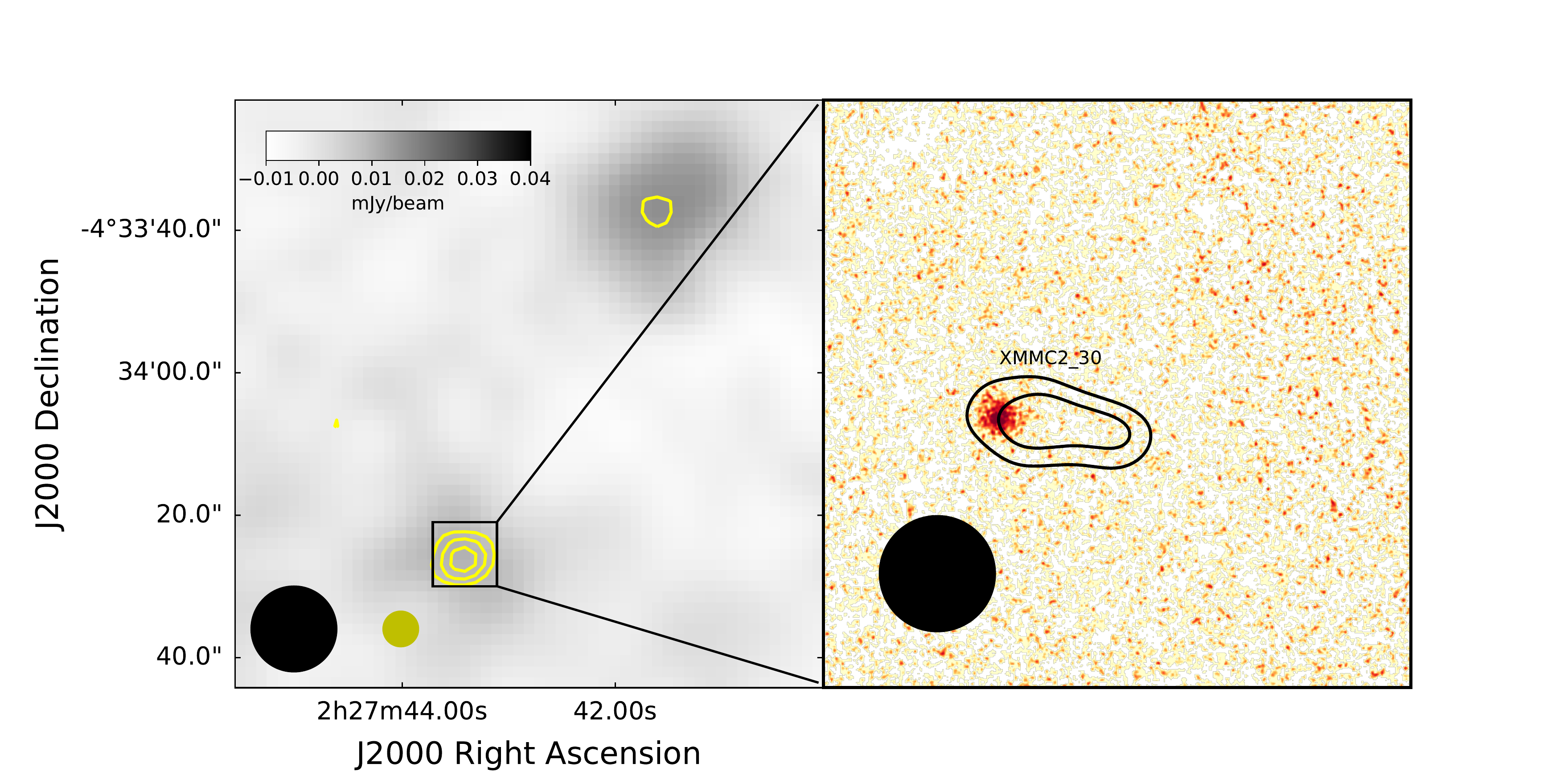}
\caption{XMMC2\_H2. The left-hand panel shows a grayscale image of the Herschel/SPIRE 250$\,\mu$m image of the XMM-C2 field with overlaid contours  of the SWIRE MIPS 24$\,\mu$m in yellow (at levels of 0.1, 0.15, 0.2, 0.3 and 0.4~MJy~sr$^{-1}$). The right-hand panel shows a zoom in of the region around XMMC2\_H2, showing the Gemini image overlaid with contours from the SWIRE 8$\mu$m image in black (contour levels 7.75 and 7.8~MJy~sr$^{-1}$). Beam sizes (FWHM) are indicated in the bottom-left corner of each image.}
\label{fig:XMMC2-H2}
\end{figure*}

\begin{figure*}[t!]
\centering
\includegraphics[clip=true, trim=2.2cm 0.5cm 1.8cm 0cm, width=5.5in]{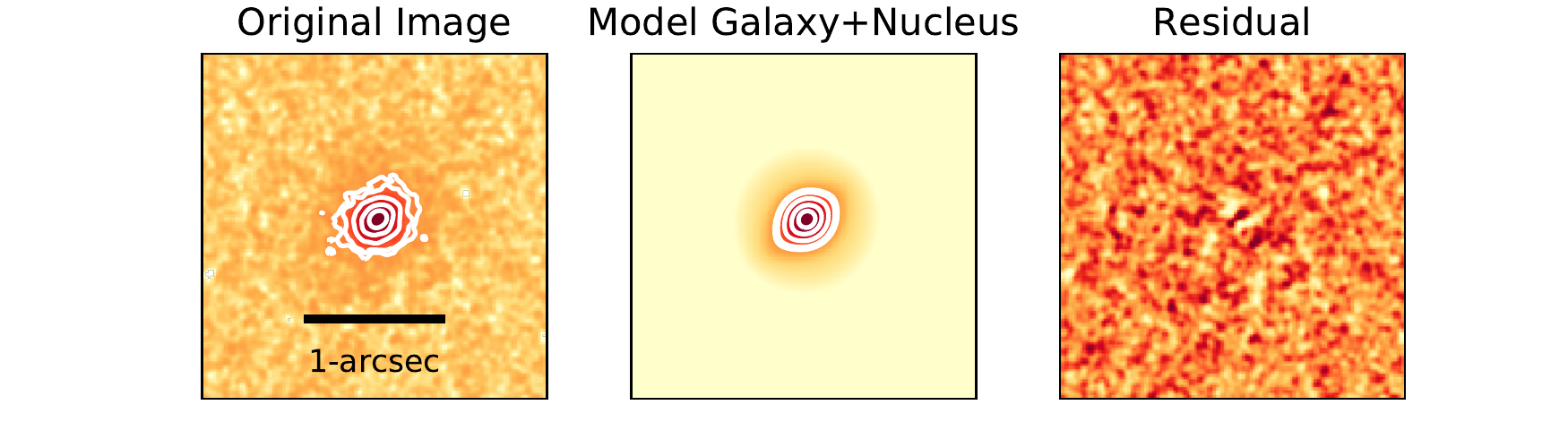}
%\centering
\caption{Tractor modeling of the Type-1 AGN ES1C\_62. Left: the original Gemini $K$-band image. Middle: the best fit model of the host galaxy (an exponential disk) plus a point source nucleus as fitted by the Tractor code.  Right: The residual obtained by subtracting the best fit model from the data. The original image and the model are shown with a logarithmic stretch and the residual image is shown with a linear stretch. Contours are at 3, 4, 8, 16 and 32 times the RMS noise.\\}
\label{fig:agn}
\end{figure*}

\subsubsection{ES1C\_H2}

Several galaxies are visible within the contours of this source (Figure \ref{fig:es1c}, top left, and Figure \ref{fig:ES1C-H2}). Two are associated with a relatively strong 24$\,\mu$m source, ES1C\_60 and ES1C\_62, and  also with weak radio sources, so we assume these are the dominant contributors to the {\em Herschel} flux.  A third source, ES1C\_55 (R.A.(J2000) $=$ 00:35:18.73, Dec.(J2000) $= -$44:00:57.3), is associated with a relatively strong radio source ($S_{\rm 8.4~GHz}=175\pm 5~\mu$Jy), but has no visible 24$\,\mu$m emission and is most likely a radio loud AGN. It has a close companion galaxy (ES1C\_56). Finally, there is also a radio source very close ($\approx$2$^{\prime \prime}$ SE) to the guide star (unfortunately too close to be able to obtain reliable optical/infrared photometry).

Both ES1C\_60 and ES1C\_62, which appear as a close pair on the sky, have SEDs consistent with a dominant AGN component. ES1C\_62 has a point source contribution approximately equal in flux to that of the extended host galaxy, so it is most likely a Type-1 (unobscured) AGN. All three AGN (ES1C\_55, ES1C\_60 and ES1C\_62) have similar photometric redshifts (1.37, 1.45 and 1.44 for ES1C\_55, ES1C\_60 and ES1C\_62, respectively, and ES1C\_56, the non-AGN companion to ES1C\_55, is at a redshift of 1.49.) Thus, this system is a candidate triple AGN (which needs to be confirmed via spectroscopy).  Only a handful of these are known (e.g.\ Liu et al.\ 2011; Schawinski et al.\ 2011; Deane et al.\ 2014). Unfortunately, its proximity to one of the guide stars will make spectroscopic confirmation of this system challenging. The guide star is 6$^{\prime \prime}$ from the nearest AGN (ES1C\_62), if scattered light can be excluded then a conventional longslit spectrum from a large ground-based telescope might be practical, but this system is probably better investigated using a space-based platform, or a spectrograph fed by adaptive optics. Even so, care will need to be taken to ensure stray light from the star does not affect the observation. 

Fitting the Composite-2 SED of Kirkpatrick et al.\ (2015) (the closest template presented in that paper, though still failing to capture some of the mid-infrared flux) suggests that the infrared emission from star formation (60\% of the total) corresponds to a star formation rate summed over the whole system of about $180\, M_{\odot}$~yr$^{-1}$ (Table \ref{tab:derived}). As star formation dominates the far-infrared flux, it is likely that this is a good estimate despite the relatively poor fit in the mid-infrared. The star formation rate estimate based on the radio emission is significantly higher, $\approx 600\,M_{\odot}$~yr$^{-1}$ (excluding the emission from the radio galaxy, ES1C\_55). However, it is likely that at least some of the radio emission is from the AGN even in the ``radio quiet'' systems.
In the HerMES catalog, this source is identified with 4HERMES~S250~SF~J003519.2-440055, from a $24\,\mu$m prior that appears to be dominated by the blend of ES1C\_60 and ES1C\_62.

We used the Tractor to decompose the light from the Type-1 AGN ES1C\_62 in the Gemini $K$-band data (Figure \ref{fig:agn}), the only AGN to show a point source nucleus. Two components were fit, a point-source nucleus representing the AGN, and an extended host galaxy. The best fit indicates the AGN and the host galaxy each contribute $\approx 18\pm 4\,\mu$Jy to the total flux of the object, with the host being best fit by a fairly compact disk galaxy ($0\farcs05 \pm 0.02$ half-light radius).  The point source flux corresponds to $M_i\approx -23.3$ at $z_{\rm phot}=1.44$. This object, and its companion Type-2 AGN host ES1C\_60 (which has a very similar luminosity and a half-light radius of $\approx 0\farcs1$), can be compared to the $z\approx 1.8$ faint quasars studied with the NICMOS instrument on {\em HST} by Ridgway et al.\ (2001). In general,
the host galaxies in the Ridgway et al.\ sample are fainter ($M_i\sim -21$ to $22$ compared to $M_i\approx -23.3$), but have larger scale sizes ($\sim 2$~kpc compared to $\approx 0.4-0.8$~kpc). Whether these differences are due to small number statistics, different host properties of infrared versus optically-selected AGN, or due to the difficulty of removing the larger {\em HST} PSF remains to be established.

% \begin{figure*}[t!]
% %\includegraphics[clip=true, trim=0.95cm 0.9cm 4.9cm 2.5cm, width=\textwidth]{ES1-C_galaxies_v4_compressed.pdf}
% %\includegraphics[scale=0.7]{agn_decomp.pdf}
% \includegraphics[clip=true, trim=1cm 0.5cm 1cm 0cm, width=\textwidth]{agn_decomp.pdf}
% %\centering
% \caption{Tractor modeling of the Type-1 AGN ES1C\_62. {\bf Left: the original Gemini $K$-band image; middle, the best fit model of the host galaxy (an exponential disk) plus a point source nucleus as fitted by the Tractor code, right: the residual obtained by subtracting the best fit model from the data. The original image and the model are shown on logarithmic scales, the right-hand image is on a linear scale. Contours are at 3,4,8,16 and 32 times the RMS noise.}}
% \label{fig:agn}
% \end{figure*}

\subsubsection{XMMC2\_H1}

This source is identified with a pair of interacting galaxies (XMMC2\_69 and XMMC2\_67; upper left inset in Figure 2) with a mean photometric redshift of $z_{\rm phot}1.5$ (1.42 and 1.60 for XMMC2\_69 and XMMC2\_67, respectively). A faint detection in the VLA data (Figure \ref{fig:XMMC2-H1}) indicates that the bulk of the $\approx 190 \, M_{\odot}\,\,{\rm yr}^{-1}$  of star formation is in XMMC2\_69. In the HerMES catalog this source is identified with 4HERMES~S250~SF~J022740.5-043322.

\subsubsection{XMMC2\_H2}

The near-infrared emission from this source at $z_{\rm phot}=2.01$ seems significantly offset from the 8 and 24$\,\mu$m peak (Figure \ref{fig:XMMC2-H2}), suggesting a highly-obscured star-forming companion that is not seen, even in the IRAC 4.5$\,\mu$m data from SERVS. The estimated star formation rate of $310 \; M_{\odot}\; {\rm yr}^{-1}$ based on the {\em Herschel} far-infrared flux is consistent with the limit from the lack of a radio detection of $<560 \,M_{\odot}\; {\rm yr}^{-1}$. In the HerMES catalog this source is identified with 4HERMES~S250~SF~J022742.4-043411.

\begin{deluxetable*}{cccccccccc}
\tablecaption{Counterparts of AGN and Herschel Sources  \label{tab:observed}}
\tablecolumns{10}
\tablewidth{0pt}
\tablehead{
\colhead{Source} & \colhead{R.A.} & \colhead{Dec.} & \colhead{$S_{5.8\,\mu{\rm m}}$} & \colhead{$S_{8.0\,\mu{\rm m}}$} & \colhead{$S_{24\,\mu{\rm m}}$} & \colhead{$S_{250\,\mu{\rm m}}$} & \colhead{$S_{350\,\mu{\rm m}}$} & \colhead{$S_{8.4\,{\rm GHz}}$} & \colhead{$q_{{\rm IR}}$} \\
\colhead{} & \colhead{(J2000)} & \colhead{(J2000)} & \colhead{($\mu$Jy)} & \colhead{($\mu$Jy)} & \colhead{($\mu$Jy)} & \colhead{(mJy)} & \colhead{(mJy)} & \colhead{($\mu$Jy)} & \colhead{} \\
\colhead{(1)} & \colhead{(2)} & \colhead{(3)} & \colhead{(4)} & \colhead{(5)} & \colhead{(6)} & \colhead{(7)} & \colhead{(8)}  & \colhead{(9)}  & \colhead{(10)}
}
\startdata
\multicolumn{10}{l}{{\bf ES1C\_H1} (4HERMES~S250~SF~J003519.2-440145 and 4HERMES~S250~SF~J003520.4-440151)} \\
ES1C\_9 & 00:35:20.79 &-44:01:55.7& $13\pm 20$ & $4\pm 20$ & $152 \pm 50$ & $5\pm 4$ & $12\pm 5$ & $9\pm 4$ & 2.7\\
ES1C\_13 & 00:35:20.11 & -44:01:50.1 & $22\pm 20$ & $132\pm 20$ & $177\pm 50$ & ($-3\pm 4$ )& ($-1\pm 5$) & $15\pm 6$ & \nodata \\
ES1C\_15 & 00:35:19.24 &-44:01:48.6 & $46\pm 20$ & $46\pm 20$ & $199\pm 50$ & $30\pm 4$ & $18\pm 5$ & $15\pm 5$&2.7\\
\hline
\multicolumn{10}{l}{{\bf ES1C\_H2} (4HERMES~S250~SF~J003519.2-440055)}\\
ES1C\_60 & 00:35:18.99 &-44:00:56.7 & $113\pm 20$ & $170\pm 20$ & $500\pm 50^a$ & $27\pm 4^a$ & $18\pm 5^a$ & $10.0\pm 2.4$&$3.1^a$\\
ES1C\_62 & 00:35:18.92 & -44:00:54.8 & $92\pm 20$ & $126\pm 20$ & \nodata & \nodata & \nodata & $10.0\pm 2.4$ & \nodata \\
\hline
\multicolumn{10}{l}{{\bf XMM2\_H1} (4HERMES~S250~SF~J022740.5-043322)}\\
XMMC2\_69 & 02:27:40.69 & -04:33:23.4 & $120\pm 20$ & $84\pm 20$ & $151\pm 50$ & $23\pm 5$ & $17\pm 8$ & $48\pm 12^b$ & 1.9\\ 
\hline
\multicolumn{10}{l}{{\bf XMM2\_H2} (4HERMES~S250~SF~J022742.4-043411)}\\
XMMC2\_30 & 02:27:43.52 & -04:34:10.9 & $54\pm 20$ & $40\pm 20$ & $277\pm 50$ & $19\pm 5$ & $11\pm 8$ & $<10^b$ & 3.0 \\ 
\enddata
\tablecomments{Column (1): Source name. Column (2): Right ascension.  Column (3): Declination.  Column (4): {\it Spitzer} 5.8$\,\mu$m flux from SWIRE. Column (5): {\it Spitzer} 8.0$\,\mu$m flux from SWIRE. Column (6): {\it Spitzer} 24$\,\mu$m flux from SWIRE. Column (7): {\it Herschel} 250$\,\mu$m flux from HERMES. Column (8): {\it Herschel} 350$\,\mu$m flux from HERMES. Column (9): 8.4~GHz flux from our new radio observations.  The data are from the ATCA unless otherwise noted.  Upper limits are given at the $3\sigma$ level. Column (10): Radio-infrared ratio.}
 \tablenotetext{a}{ES1C\_60 and ES1C\_62 have a separation of $\sim2^{\prime \prime}$, and are thus too close to be deblended in the 24, 250 and 350$\,\mu$m bands. Thus, the infrared fluxes and $q_{\rm IR}$ correspond to the sum of the two objects.}
\tablenotetext{b}{Calculated based on the $8-12$~GHz VLA flux density and converted to 8.4~GHz assuming a radio spectral index of $\alpha = -0.7$.}
\end{deluxetable*}
\begin{deluxetable*}{ccccccccccc}
\tablecaption{Derived properties of {\em Herschel} sources  \label{tab:derived}}
\tablecolumns{9}
\tablewidth{0pt}
\tablehead{
\colhead{Source} & \colhead{$z_{\rm phot}$} & \colhead{$L_{\rm IR}$} & \colhead{SFR (IR)} & \colhead{$L_{1.4\,{\rm GHz}}$} & \colhead{SFR (radio)} & \colhead{$r_{1/2}$} & \colhead{$M_*$} & \colhead{Model}\\
\colhead{} & \colhead{} & \colhead{($10^{12} L_{\odot}$)} & \colhead{($M_{\odot}$~yr$^{-1}$)} & \colhead{(W~Hz$^{-1}$)} & \colhead{($M_{\odot}$~yr$^{-1}$)} & \colhead{(kpc)} & \colhead{($M_{\odot}$)} & \colhead{} \\
\colhead{(1)} & \colhead{(2)} & \colhead{(3)} & \colhead{(4)} & \colhead{(5)} & \colhead{(6)} & \colhead{(7)} & \colhead{(8)} & \colhead{(9)}   
}
\startdata
{\bf ES1C\_H1} & & & & & & \\
ES1C\_9 &3.0 & $4.6 $  & 790 & 2.6$\times$10$^{24}$ & 1400 & 1.7 & $1.4\times 10^{11}$ & Exp\\
ES1C\_15 & 2.33 & $5.0$ & 860 & 2.4$\times$10$^{24}$ & 1300 & 3.8 & $3.1\times 10^{11}$&Exp\\
\hline
{\bf ES1C\_H2} & & 2.6 & 180$^{\dag}$ & & & & & \\
ES1C\_55 & 1.37 & \nodata & \nodata & 9.2$\times$10$^{24}$ & \nodata & 1.8& $9.5\times 10^{10}$&Dev\\
ES1C\_60 & 1.45 & \nodata & \nodata & 5.2$\times$10$^{23}$ & 290 & 0.9 & $8.1\times 10^{10}$&Exp\\
ES1C\_62 & 1.44 & \nodata & \nodata & 5.2$\times$10$^{23}$ & 290 & 0.4 & $1.2\times 10^{11}$&Exp\\ 
\hline
{\bf XMM2\_H1} & & & & & \\
XMMC2\_69 & 1.61 & $1.1$ & 190 & 2.9$\times$10$^{24}$ & 1600 & 10.5 & $7.9\times 10^{10}$&Dev \\
\hline
{\bf XMM2\_H2} & & & & \\
XMMC2\_30 & 2.01 & $1.8$ & 310 & $<$1.0$\times$10$^{24}$ & $<$560 & 1.9 & $1.2\times 10^{11}$&Exp\\
\enddata
\tablecomments{$\dag$ $L_{IR}$ and SFR(IR) for ESC1\_H2 are for the sum of the IR-bright galaxies ES1C\_60 and ES1C\_62.
Column (1): Source name.  Column (2): Photometric redshift.  Column (3): Infrared luminosity. Column (4): Star formation rate estimated using the far-infrared luminosity. Column (5): 1.4~GHz radio luminosity from VLA or ATCA observations (assuming a spectral index of -0.8).  Column (6): Radio star formation rate (Section 3.8). Column (7): Half-light radius in kpc from Tractor fitting.  Column (8): Stella mass.  Column (9): Best fitting surface brightness profile model (Exp = Exponential; Dev = de Vaucouleurs) from the Tractor photometry.
}
\end{deluxetable*}
%%%%%%%%%%%%%%%%%%%%%%%%%%%%%%%%%%%%%%%%%%%%%%%%

\subsection{Other individual objects}\label{sec:other}
We briefly mention some other interesting galaxies highlighted in Figures \ref{fig:es1c}, \ref{fig:cdfsc} and \ref{fig:xmmc2}, and listed in Table \ref{tab:others}. ES1C\_46 is a disturbed disk galaxy with $z_{\rm phot}=0.9$; it is weakly detected at $24\,\mu$m. The CDFS-C field has the best image quality of our fields, but unfortunately lacks {\em Herschel} detections. Nevertheless, we identify several high redshift galaxies (Figure \ref{fig:cdfsc}). CDFSC\_60 is a compact, $z_{\rm phot}=3.2$ galaxy with a half-light radius of $0\farcs032$ (0.24~kpc), and CDFSC\_11 and CDFSC\_12 are an apparent pair of galaxies at $z_{\rm phot}=2.9$.

% \begin{table}
% \caption{Other objects of interest}
% \begin{tabular}{lcccl}
% Name & $z_{\rm phot}$ & $M_*$ & Radius & Tractor \\
%            &                         &        ($M_{\odot}$)              &        Model\\\hline\hline
% ES1C\_46 & 0.9 & $2.5\times 10^{10}$&3.8 & Exp\\ 
% CDFSC\_60 & 3.2 & $2.0\times 10^{10}$& 0.24 &Dev \\
% CDFSC\_11 & 2.9 &$1.3\times 10^{10}$ & 1.1 &Exp \\
% CDFSC\_12 & 2.9 &$4.0\times 10^{10}$  & 0.5 & Exp\\\hline
% \end{tabular}
% \label{tab:others}
% \end{table}

%%%%%%%%%%%%%%%%%%%%%%%%%%%%%%%%%%%%%%%%%%%%%%%
\begin{deluxetable}{ccccc}
\tablecaption{Other Objects of Interest \label{tab:others}}
\tablecolumns{4}
\tablewidth{0pt}
\tablehead{
\colhead{Source} & \colhead{$z_{\rm phot}$} & \colhead{$M_*$} & \colhead{$r_{1/2}$} & \colhead{Model}  \\
\colhead{} & \colhead{} & \colhead{($M_{\odot}$)} & \colhead{(kpc)} & \colhead{} \\
\colhead{(1)} & \colhead{(2)} & \colhead{(3)} & \colhead{(4)} & \colhead{(5)}
}
\startdata
ES1C\_46 & 0.9 & $2.5\times 10^{10}$&3.8 & Exp\\ 
CDFSC\_60 & 3.2 & $2.0\times 10^{10}$& 0.24 &Dev \\
CDFSC\_11 & 2.9 &$1.3\times 10^{10}$ & 1.1 &Exp \\
CDFSC\_12 & 2.9 &$4.0\times 10^{10}$  & 0.5 & Exp\\
\enddata
\tablecomments{Column (1): Source name.  Column (2): Photometric redshift.  Column (3): Stellar mass.  Column (4): Half-light radius in kpc from the Tractor fitting. Column (5): Best fitting surface brightness profile model (Exp = Exponential; Dev = de Vaucouleurs) from the Tractor photometry.}
\end{deluxetable}
%%%%%%%%%%%%%%%%%%%%%%%%%%%%%%%%%%%%%%%%%%%%%%%%

\section{Discussion}

%This pilot survey has shown the potential for MCAO applications for observations of $z\sim 1-4$ galaxies. 
We have presented some of the highest angular resolution images of $z\stackrel{>}{_{\sim}}1$ galaxies obtained in the near-infrared. Our results on the evolution of galaxy sizes are broadly consistent with larger studies made with the {\em HST} (van der Wel et al.\ 2014, Shibuya et al.\ 2015, Allen et al.\ 2017), though we see some differences, particularly in the size distribution of moderately massive $>10^{10}\, M_{\odot}$ star-forming galaxies at  $z>2$, where we see a higher fraction of compact (radius $<2$~kpc) objects ($\approx$90\% of the population in this study compared to $\approx 60$\% with {\em HST}) that will need further Gemini data, or observations with the {\em JWST} to resolve. 

The launch of {\em JWST} will enable studies of large numbers of the field galaxy population at similar angular resolution and very high sensitivity, but the difficulty of performing large surveys with {\em JWST} will leave a niche for ground-based MCAO to make targeted observations of large numbers of rare objects. We have therefore emphasized the study of the hosts of ULIRGs in our fields. 
By combining the Gemini data with  infrared data from {\em Spitzer} and {\em Herschel}, together with arcsecond or better resolution radio data, we show that most of the {\em Herschel} sources appear to arise from multiple systems, including one candidate triple AGN that would not have been otherwise identified as a multiple system\footnote{Although the triple AGN system is resolved in ground-based imaging, it is cataloged as a single detection in the VIDEO near-infrared catalog.}. The high resolution of GeMS at wavelengths in the rest frame optical is thus an essential need for accurately classifying these systems. For example, Capelo et al.\ (2016) use simulations to estimate that $\sim 20$\% of AGN are dual, but that high resolution ($\stackrel{<}{_{\sim}} 10~$kpc) data are needed to identify them. Observations such as these can therefore significantly improve the constraints on 
%models for dual AGN activity. 
galaxy and supermassive black hole merger rates.  

Figure \ref{fig:sizes} compares the counterparts of the {\em Herschel} sources to field galaxies at similar redshifts. As expected, they tend to be the more massive galaxies. The corollary of this is that a large fraction, $\stackrel{>}{_{\sim}}50$\% of the $\stackrel{>}{_{\sim}}10^{11}\,M_{\odot}$ galaxies at $z>1$ in our sample, show strong AGN or starburst activity, also seen in other samples of massive high-$z$ galaxies (Spitler et al.\ 2014; Martis et al.\ 2016; Schreiber et al.\ 2017).  Also as expected, the {\em Herschel} galaxies fall on the red side of the star-forming population, and the radio galaxy (ES1C\_55) plots amongst the quiescent objects. 
The AGN hosts (ES1C\_55,60 and 62) also tend to have more compact scale sizes than the star-forming galaxies, a trend that has also been noted for high redshift AGN selected from WISE (Farrah et al.\ 2017).

Previous studies of {\em Herschel} and other high-redshift, submillimeter-selected galaxies find broadly similar results. Kartaltepe et al.\ (2012) examined the morphologies of 52 $0.5<z<2$ ULIRGs selected from the GOODS-S field at 100 and 160$\,\mu$m (using PACS photometry), finding that the merger fraction in {\em HST} F160W images was higher in the ULIRGs (50-73\%) compared to a matched sample of field galaxies on the star forming ``main sequence" of star formation rate vs stellar mass (24\%).  With only a few objects it is difficult to estimate a merger rate, but of the four {\em Herschel} sources we have examined, we have one close pair (XMM-C2\_69), one candidate triple AGN (ES1C\_60, 62, and 55), and two objects at different redshifts associated with the same {\em Herschel} source with disturbed morphologies that might be late stage mergers (ES1C\_9 and ES1C\_15). Thus, we are certainly consistent with the results of Kataltepe et al.\ (2012). In contrast, Targett et al.\ (2013) found few mergers in their {\em HST}/WFC3 study of $1<z<3$ galaxies selected in the submillimeter, but did find, as we do, that the $z>2$ star-forming ULIRG  population is dominated by large disk galaxies (in contrast to the AGN in this sample, which have a mix of profiles and smaller half-light radii). 

The {\em Herschel} systems studied in this paper can also be compared to the high resolution studies of $z\sim 4$ star-forming galaxies by Wiklind et al.\ (2014), Hodge et al. (2016) and Rujopakarn et al.\ (2016) with {\em HST} and radio/submillimeter imaging with the VLA and ALMA, which reveal similar trends, namely disturbed host galaxy morphologies and star-forming regions offset from the centroid of the stellar light (XMMC2\_30 is a very good example). The fact that these morphologies persist into the K-band (rest-frame optical) in the sample studied in this paper suggest that the appearance of these objects is not dominated by clumpy extinction, but instead reflects the actual distribution of stellar light. The {\em Herschel}/SPIRE selected objects in this paper thus provide 
%an interesting bridge 
a bridge between objects discovered in deep {\em Herschel}/PACS observations at shorter wavelengths and mostly lower redshifts, and the more extreme objects (both in terms of luminosity and redshift) discovered in the submillimeter surveys.

%A planned upgrade for GeMS will allow it to be used with guide stars as faint as $R\approx 17$. When enacted, this will open up a much larger area of the extragalactic sky to this type of observation. Although ground-based images will not be able to compete with space-based images from {\em JWST} in terms of depth, the ability to survey a {\bf large number of targets} much more efficiently than {\em JWST}  will make GSAOI a useful tool for following up large numbers of rare objects such as submm galaxies and the hosts of luminous AGN/quasars. With a larger sample, we can test whether the trends suggested in this pilot study are present in the population as a whole.

%\section{Summary}
%{\bf \color{blue}KN is working on this . . . }

A planned upgrade for GeMS will allow it to be used with guide stars as faint as $R\approx 17$. When enacted, this will open up a much larger area of the extragalactic sky to subarcsecond-resolution near-infrared observations. Although ground-based imaging will not be able to compete with space-based observations from {\em JWST} in terms of depth, the ability to survey a large number of widely-separated targets much more efficiently than {\em JWST}, and with a much less oversubscribed facility will make GSAOI a useful tool for following up large numbers of rare objects. These might include luminous submillimeter galaxies and the hosts of luminous AGN/quasars. With a larger sample, we can test whether the trends suggested in this pilot study are present in the population as a whole.

\acknowledgements
We thank J.C.\ Mauduit for assistance with the CTIO observations in this paper. 
J.A. gratefully acknowledges support from the Science and Technology Foundation (FCT, Portugal) through the research grant UID/FIS/04434/2013.
The National Radio Astronomy Observatory is a facility of the National Science 
Foundation operated under cooperative agreement by Associated Universities, Inc.
This work made extensive use of {\sc topcat} (Taylor 2005) for catalog matching
and analysis, and the Virtual Observatory SAMP protocol for communication
between applications. This research made use of Montage. It is funded by the National Science Foundation under Grant Number ACI-1440620, and was previously funded by the National Aeronautics and Space Administration's Earth Science Technology Office, Computation Technologies Project, under Cooperative Agreement Number NCC5-626 between NASA and the California Institute of Technology. This research made use of APLpy, an open-source plotting package for Python (Robitaille and Bressert, 2012).

\end{document}